\documentclass[sigconf,screen]{acmart}

\newcommand{\sys}{\mbox{\textsc{Ganadi}}\xspace}
\newcommand{\cent}{\mbox{\textsc{Centris}}\xspace}
\newcommand{\tplite}{\mbox{\textsc{TPLite}}\xspace}

\newcommand{\vzfinder}{\mbox{\textsc{V0Finder}}\xspace}

\newcommand{\cn}{\mbox{\textsc{Cneps}}\xspace}

\usepackage{xstring}
\newcommand{\PP}[1]{
	\vspace{4px}
\noindent{\bf \IfEndWith{#1}{.}{#1}{#1.}}}

\newcommand{\PPP}[1]{
	\vspace{4px}
{\textit{\textbf{\IfEndWith{#1}{.}{#1}{#1.}}}}}

\newcommand{\PI}[1]{
\vspace{4px}
\textit{\textbf{\IfEndWith{#1}{.}{#1}{#1.}}}}

\newcommand{\etal}{\textit{et al}.\xspace}
\newcommand{\ie}{\textit{i}.\textit{e}.}
\newcommand{\eg}{\textit{e}.\textit{g}.}

\usepackage{algorithmic}
\usepackage{graphicx}
\usepackage{xspace}
\usepackage{xcolor}
\def\BibTeX{{\rm B\kern-.05em{\sc i\kern-.025em b}\kern-.08em
    T\kern-.1667em\lower.7ex\hbox{E}\kern-.125emX}}

\usepackage{float}
\newfloat{lstfloat}{t}{lop}
\floatname{lstfloat}{Listing}

\newcommand{\algorithmautorefname}{Algorithm}
\newcommand{\refappendix}[1]{\hyperref[#1]{Appendix~\ref*{#1}}}
\def\sectionautorefname~#1\null{Section #1\null}
\def\equationautorefname~#1\null{Equation (#1)\null}
\def\algorithmautorefname~#1\null{Algorithm (#1)\null}
\def\subsectionautorefname~#1\null{Section #1\null}
\def\subsubsectionautorefname~#1\null{Section #1\null}

\usepackage{arydshln}
\usepackage{kotex}
\usepackage{tikz}
\usepackage{url}
\usepackage{microtype}
\usepackage{lipsum}
\usetikzlibrary{positioning}

\usepackage[skins]{tcolorbox}
\tcbuselibrary{breakable}
\definecolor{titlecolor}{rgb}{0.9490196078431373, 0.8627450980392157, 0.796078431372549}
\definecolor{oorange}{rgb}{0.937, 0.396, 0.282}
\definecolor{yyellow}{rgb}{0.992, 0.733, 0.517}
\definecolor{bbage}{rgb}{0.996, 0.909, 0.784}

\usepackage{booktabs}
\PassOptionsToPackage{hyphens}{url}
\makeatletter
\g@addto@macro{\UrlBreaks}{\UrlOrds}
\makeatother
\definecolor{blue(ryb)}{rgb}{0.01, 0.28, 1.0}
\definecolor{oceanboatblue}{rgb}{0.0, 0.47, 0.75}
\definecolor{navyblue}{rgb}{0.05, 0.0, 0.61}
\definecolor{tomatoorig}{rgb}{0.7, 0.0, 0.0}
\definecolor{mediumelectricblue}{rgb}{0.01, 0.31, 0.59}

\definecolor{minusgray}{rgb}{0.9, 0.9, 0.9}
\definecolor{citeyellow}{rgb}{0.56, 0.46, 0}
\definecolor{plusgray}{rgb}{1, 0.91, 0.5}

\definecolor{mygray}{rgb}{0.4,0.4,0.4}
\definecolor{myblue}{rgb}{0.1, 0, 0.58}
\definecolor{gostop}{rgb}{0, 0.56, 0.35}

\definecolor{skyblue}{rgb}{0.82, 0.91, 1}
\definecolor{mygreen}{rgb}{0.08, 0.33, 0.20}
\definecolor{tomato}{rgb}{0.7, 0.0, 0.0}

\definecolor{asered}{rgb}{0.690196078, 0, 0.250980392}
\definecolor{asegreen}{rgb}{0, 0.501961, 0}
\definecolor{aseblue}{rgb}{0.2156862745098039, 0.2156862745098039, 1}

\definecolor{mypink}{rgb}{1, 0.9, 0.91}
\definecolor{mypink}{rgb}{0.98, 0.925, 0.929}
\definecolor{mydiffgreen}{rgb}{0.855, 0.96, 0.886}

\definecolor{newgreen}{rgb}{0.133333333, 0.694117647, 0.298039216}
\definecolor{mynavy}{rgb}{0.15, 0.15, 0.35}

\definecolor{vudch}{rgb}{0.1, 0.66, 0.1}
\definecolor{greentwo}{rgb}{0.06, 0.81, 0.06}
\definecolor{icsepink}{rgb}{0.95, 0.69, 0.75}
\definecolor{icseorange}{rgb}{0.98, 0.52, 0.36}

\usepackage{mdframed}

\definecolor{mycyan}{rgb}{0.5, 1, 0.9}
\definecolor{myorange}{rgb}{1, 0.6, 0}
\definecolor{algreen}{rgb}{0.63, 0.83, 0.41}
\definecolor{lightgray}{rgb}{0.8, 0.8, 0.8}
\definecolor{citegray}{rgb}{0.4, 0.4, 0.4}

\usepackage{hhline}

\usepackage{subcaption}
\usepackage{listings}

\usepackage{ragged2e}
\usepackage{multirow}
\usepackage{breakurl}

\definecolor{algreen}{rgb}{0, 0.5,0.25}

\usepackage[ruled, vlined, linesnumbered]{algorithm2e}
\SetKwInOut{Parameter}{parameter}
\SetCommentSty{mycommfont}
\SetAlgorithmName{Algorithm}{Algorithm}
\SetAlCapNameFnt{\footnotesize}
\SetAlCapFnt{\footnotesize}
\usepackage{color,soul}
\usepackage[skins]{tcolorbox}
\tcbuselibrary{breakable}

\lstdefinestyle{base}{
	language=C,
	numbers=left,
	numberstyle=\tiny,
	numbersep=-3pt,
	breaklines=true,
	commentstyle=\color{mygray},
	columns=fullflexible,
	basicstyle=\fontsize{9}{9}\ttfamily\color{black},
	moredelim=**[is][\bfseries]{@bold-}{-bold@},
	moredelim=**[is][\color{tomato}]{@R-}{-R@},
	moredelim=**[is][\color{mygreen}]{@G-}{-G@},
	moredelim=**[is][\color{mygray}\bfseries]{@B-}{-B@},
	moredelim=**[is][\color{blue}\bfseries]{@AB-}{-AB@},
	escapeinside={(**}{**)},
	tabsize=4,
	xleftmargin=1pt,
	captionpos=t,
	frame=bt,
	framesep=2pt,
	framerule=0.5pt,
	showstringspaces=false
}

\makeatletter
\newcommand{\thickhline}{%
	\noalign {\ifnum 0=`}\fi \hrule height 1.5pt
	\futurelet \reserved@a \@xhline
}
\makeatother

\AtBeginDocument{%
  \providecommand\BibTeX{{%
    Bib\TeX}}}

\copyrightyear{2026}
\acmYear{2026}
\setcopyright{cc}
\setcctype{by}
\acmConference[ASE '26]{Proceedings of the 41st IEEE/ACM International Conference on Automated Software Engineering}{October 12--16, 2026}{Munich, Germany}
\acmBooktitle{Proceedings of the 41st IEEE/ACM International Conference on Automated Software Engineering (ASE '26), October 12--16, 2026, Munich, Germany}
\acmDOI{10.1145/3832783.3837498}
\acmISBN{979-8-4007-2882-2/2026/10}
\acmSubmissionID{ase26main-p1485-p}
\received{2026-03-26}
\received[accepted]{2026-06-18}

\begin{document}

\title{GANADI: Uncovering C/C++ OSS Reuse Genealogies via Pivotal Function-Based Clustering to Enhance Supply Chain Security}

\author{Dongyeon Kim}
\orcid{0009-0001-7283-5547}
\affiliation{%
  \institution{Korea University}
  \city{Seoul}
  \country{Republic of Korea}
}
\email{kimdongyeon@korea.ac.kr}

\author{Seunghoon Woo}
\correspondingauthor
\orcid{0000-0002-5455-0804}
\affiliation{%
  \institution{Korea University}
  \city{Seoul}
  \country{Republic of Korea}
}
\email{seunghoonwoo@korea.ac.kr}

\author{Heejo Lee}
\correspondingauthor
\orcid{0000-0002-5831-0787}
\affiliation{%
  \institution{Korea University}
  \city{Seoul}
  \country{Republic of Korea}
}
\email{heejo@korea.ac.kr}

\begin{abstract}
We present \sys, a systematic approach for identifying C/C++ OSS 
reuse genealogies to enhance software supply 
chain security.
Understanding OSS reuse genealogy is crucial for improving SBOM 
completeness 
and prioritizing 
security remediation across supply chains.
Although existing approaches can identify reused components and 
vulnerabilities within a project, they fail to trace 
OSS reuse paths through intermediate projects, limiting their 
effectiveness in securing supply chain ecosystems.
To address this limitation, 
\sys constructs reuse genealogies 
by clustering downstream projects based on shared characteristics of origin-derived code (called pivotal functions), and then inferring reuse direction among the 
projects within each cluster.
When applied to 20 widely reused OSS projects with over 1,500 propagation 
paths, \sys achieved 84.85\% precision 
and 95.76\% recall in identifying reuse genealogies,
outperforming existing approaches that achieved at most 23.21\% recall.
Leveraging OSS reuse genealogy for vulnerability detection, 
we identified 48 unpatched vulnerabilities in real-world popular C/C++ projects.
Among them, 23 were patched following our responsible disclosure (including one CVE ID assigned), demonstrating the practical impact of genealogy-based vulnerability management.
\end{abstract}



\maketitle

\section{Introduction}

Open-source software (OSS) is widely reused in modern software development,
accelerating development cycles and reducing implementation costs. 
However, this widespread reuse poses security risks that are often overlooked by existing approaches. 

When developers copy-paste, fork, or adapt OSS components, they create hidden propagation paths through which vulnerabilities can spread across projects, yet these reuse relationships are rarely documented or tracked. 

Although a Software Bill of Materials (SBOM) documents which OSS components are used in a project~\cite{camp2021sbom, o2023impacts, williams2025research}, prior SBOM-based approaches do not fully capture supply chain security. Effective security analysis requires understanding not just \textit{what} components are used, but \textit{how} they are propagated: through which intermediate projects and with what modifications. Accordingly, recent SBOM standard updates~\cite{cisa2025sbom} have strengthened requirements for documenting dependency relationships.

We address this gap by reconstructing the \textit{reuse genealogy}: 
the origin-to-descendant relationships that reveal how code transitively flows through the OSS ecosystem. 
This not only enhances SBOM completeness,
but also enables understanding the propagation paths and scope of vulnerabilities in complex supply chain ecosystems, thereby facilitating more effective vulnerability management (\autoref{subsec:app}).
For example, if a project reuses
the \texttt{zlib} compression library, it is essential to determine whether the code was reused directly from the origin \texttt{zlib} or through intermediate projects
that may have modified it. Tracing this path is critical for security
management, as it reveals how vulnerabilities spread~\cite{woo2021v0finder, liu2022demystifying, huang2024vmud},
how licenses are modified~\cite{di2010exploratory, vendome2015and, wu2024large}, and how code changes accumulate across intermediate projects (\autoref{subsec:moti_ex}).

To our knowledge, no previous work has focused on identifying C/C++ OSS reuse genealogies.
Even when developing new techniques, the problem remains non-trivial due to two technical challenges (see \autoref{subsec:problem}).
First, \textit{lack of explicit reuse metadata}. C/C++ projects typically reuse code through direct copying without declaring dependencies or maintaining reuse records. Consequently, there are no explicit indicators to identify what code was reused or how it propagated~\cite{woo2021centris, jiang2023third}.
Second, \textit{difficulty in determining reuse direction}.
Even when available metadata is used to infer reuse genealogy, determining the direction of reuse remains challenging. Common metadata features, such as commit timestamps, are often inconsistent and easily manipulated during reuse~\cite{steidl2014incremental, woo2021v0finder, woo2022movery}.

\vspace{0.1em}
\PP{Limitation of existing approaches}
Existing software composition analysis (SCA) approaches (\eg, \cite{woo2021centris, jiang2023third, wu2023ossfp, duan2017identifying}) 
have focused on identifying reused components and inferring their origins, without considering the reuse genealogy.
For example, \cent~\cite{woo2021centris}, \tplite~\cite{jiang2023third}, and \textsc{BinaryAI}~\cite{jiang2024binaryai} attempt to identify components using function birth time, metadata, and function similarity, respectively. 
Because they fail to capture the directionality of reuse, they cannot resolve issues that emerge from the absence of reuse genealogy analysis, such as hidden vulnerability propagation. 
Existing propagated vulnerability detection approaches (\eg, \cite{kim2017vuddy, woo2022movery, woo2023v1scan, feng2024fire}) 
cannot track transitive vulnerability propagation across the broader supply chain or systematically identify all affected downstream projects, limiting their applicability to supply chain-wide security management.
Several studies attempt to track software evolution (\eg,\cite{inoue2012does,nguyen2013study, godfrey2008past}), but they focus on changes within a single project; thus, they do not capture reuse genealogy across multiple projects.

To address these limitations, we present \sys (Genealogy ANAlysis for Dependency Inference), a new approach for identifying OSS reuse genealogies. \sys uses \textit{pivotal functions} (\ie, functions in downstream files that contain at least one function identical to one in the origin OSS) to focus on reused code and exclude irrelevant code. It then performs (1) \textit{software clustering}, (2) \textit{reuse inference}, and (3) \textit{graph construction}.

\PP{Approach overview} 
\sys begins by constructing a pool of popular C/C++ software projects. 
Given an OSS (origin), 
\sys identifies candidates within the pool that share identical functions with the origin.
For each candidate, \sys identifies pivotal functions from files containing origin-derived code (\autoref{subsubsec:pivotal}).
It then performs software clustering of candidates based on shared characteristics (\eg, file paths). This effectively groups related projects and filters out unrelated software (\autoref{subsubsec:clustering}).

Within each group, \sys examines pairwise reuse relationships and aggregates them to construct the consolidated genealogy (\autoref{subsec:p2}).
Here, \sys identifies reuse relationships by jointly analyzing code similarity and reuse directionality, leveraging both explicit (\eg, fork) and implicit evidence (\eg, code birth dates).
Finally, \sys constructs a \textit{reuse genealogy graph} (\autoref{subsec:p3})
by connecting identified reuse relationships from all clusters to the origin, following priority-based rules to minimize false alarms.

\PP{Evaluation}
We constructed a software pool of 2,500 popular C/C++ projects from GitHub, and applied \sys to 20 widely reused OSS projects as origins to identify their reuse genealogies within the pool. \sys identified over 1,500 reuse relationships with 84.85\% precision and 95.76\% recall, outperforming existing approaches that achieved at most 23.21\% recall due to their limited ability to capture transitive reuse propagation (\autoref{subsec:accuracy}).
To evaluate \sys from a supply-chain security perspective, we incorporated reuse genealogy into a vulnerability management process. 
Across 20 widely reused OSS projects, incorporating genealogy into existing detectors~\cite{kim2017vuddy, feng2024fire} improved precision by 13.7\%, maintained comparable F1-scores using only 11\% of the baseline dataset, and clarified vulnerability propagation during manual analysis (\autoref{subsec:app}).
Using \sys, we identified 48 unpatched vulnerabilities in real-world projects, 23 of which were patched following our disclosure.

\PP{Contributions}
We summarize our contributions below.

\begin{itemize}
    \setlength\itemsep{0.12em}
    \item 
    We identify the importance of OSS reuse genealogy analysis, and propose \sys, a novel approach to identify C/C++ OSS reuse genealogies to enhance supply chain security.
    \item 
    We introduce a pivotal function-based clustering approach that enables accurate reuse relationship identification even in challenging scenarios with many intermediate projects.
    \item \sys identifies reuse genealogies with 84.85\% precision and 95.76\% recall, outperforming prior approaches, and uncovers 48 unpatched vulnerabilities through genealogy-driven vulnerability discovery.
\end{itemize}
\section{Motivation}\label{sec:moti}
We introduce basic terms and the problem with its technical challenges, then motivate our approach through a concrete example.

\subsection{Basic Terms}

We define three basic terms: OSS components, OSS reuse, and reuse relationships.
\textit{OSS components} represent a whole or part of OSS that can be independently reused in other software~\cite{woo2021centris, na2024cneps}. 
\textit{OSS reuse} refers to the utilization of OSS components within the target software.
\textit{reuse relationship} refers to the link between two software projects in which one project reuses code from the other.

\subsection{Problem Overview and Challenges}\label{subsec:problem}

Recent software projects are composed of a mix of proprietary code and reused components~\cite{woo2021centris, jiang2023third}.  
Let \mbox{$F(X) = C(X)\cup D(X)$} denote the set of functions in project $X$, where 
$C(X)$ is the set of reused functions and $D(X)$ is the set of self-developed functions.

Given a target software $X$, the traditional goal of SCA approaches (\eg, \cite{woo2021centris, wu2023ossfp, jiang2024binaryai}) is to identify $C(X)$ and the corresponding \textit{origin projects} from which each function in $C(X)$ was reused.

In contrast, 
\sys aims to identify all software projects that have reused a given origin project $o$ (directly or transitively), as well as the propagation paths of this reuse. Let $\mathcal{R}(o)$ denote the set of all projects that reused $o$, where $P$ denotes a software project:
$\mathcal{R}(o) = \{P \mid C(P) \text{ contains functions derived from } o\}$.

Our goal is to identify the reuse genealogy by constructing a directed acyclic graph $G_o = (V, E)$ where  $V = \mathcal{R}(o) \cup {\{o\}}$ and $E = \{(X, Y) \mid Y \text{ directly reuses code from } X\}$.

\vspace{0.1em}
\PP{Technical challenges}
However, addressing this problem is a non-trivial task
mainly due to the following two challenges.
First, \textit{the lack of explicit linkage in C/C++ code reuse}. Although package managers for C/C++ such as \texttt{Conan} exist,
code reuse primarily occurs via direct copy-and-paste~\cite{woo2021centris}.
As a result, reused components do not leave clear structural or metadata traces, making it difficult to detect reuse relationships through conventional static or dependency analysis. 

Second, \textit{the difficulty of determining reuse direction}.
Even when relying on timestamp-based metadata such as commit dates~\cite{woo2021v0finder}, inferring the direction or path of reuse remains challenging, especially when the origin is reused in parallel by multiple projects.
For example, if project $A$ is reused by both $B$ and $D$, which are later reused by 
$C$ and $E$, respectively, the two reuse paths 
$A\rightarrow B \rightarrow C$ and $A\rightarrow D \rightarrow E$ become interleaved in time.
In such cases, the fact that $B$'s reuse occurred before $D$'s does not imply a reuse path from 
$B$ to $D$.
Other features (\eg, commit messages or file structures) are often noisy, inconsistent, or missing in the OSS ecosystem~\cite{steidl2014incremental, woo2021v0finder}.

\vspace{0.2em}
\subsection{Motivating Example}\label{subsec:moti_ex}
\vspace{0.1em}

Without a clear understanding of OSS reuse genealogy, it is challenging to address security issues throughout the software supply chain.
Suppose we identify the OSS components used in \texttt{Redis} and \texttt{Dragonfly}, and observe that both include \texttt{Lua}.
Existing SCA techniques, which rely on component names and versions, attempt to detect known vulnerabilities by checking whether the version of \texttt{Lua} used in each project is affected~\cite{woo2021centris, woo2023v1scan,kwon2021octopocs}, referencing public vulnerability databases (\eg, NVD).

\begin{figure}[t]
\begin{lstlisting}[caption={\label{lst:redis}A patch snippet for CVE-2020-14147 (\texttt{ef764d}). Neither this code nor the patch exists in the upstream \texttt{Lua}.},   
    language=C, frame=single, style=base, frame = none,         
    basicstyle=\fontsize{7}{8}\ttfamily\color{black}, mathescape, breakatwhitespace=true, breaklines=true]
$\;\;\;\text{\textcolor{gray}{//Path: Redis/deps/}}\text{\textcolor{black}{\textbf{\underline{lua}}}}\text{\textcolor{gray}{/src/lua\_struct.c}}$
$\;\;\;\text{\textcolor{purple}{-\;static int getnum (const char **fmt, int df) \{}}$
$\;\;\;\text{\textcolor{asegreen}{+\;static int getnum (lua\_State *L, const char **fmt, int df) \{}}$
$\;\;\;\;\;\;\;\text{\textcolor{black}{...}}$
$\;\;\;\;\;\;\;\text{\textcolor{black}{int a = 0;}}$
$\;\;\;\;\;\;\;\text{\textcolor{black}{do \{}}$
$\;\;\;\text{\textcolor{asegreen}{+}}\;\;\;\;\;\;\text{\textcolor{asegreen}{if (a > (INT\_MAX / 10) || a * 10 > (INT\_MAX - (**fmt - `0')))}}$
$\;\;\;\text{\textcolor{asegreen}{+}}\;\;\;\;\;\;\;\;\;\text{\textcolor{asegreen}{luaL\_error(L, ``integral size overflow'');}}$
$\;\;\;\;\;\;\;\;\;\;\;\text{\textcolor{black}{a = a*10 + *((*fmt)++) - `0';}}$
\end{lstlisting}
\end{figure}

\begin{figure}[t]
\begin{lstlisting}[caption={\label{lst:dragon}A vulnerable code snippet found in the modified \texttt{Lua} within \texttt{Redis} reused by \texttt{Dragonfly} (now patched following our disclosure).},   
    language=C, frame=single, style=base, frame = none,         
    basicstyle=\fontsize{7}{8}\ttfamily\color{black}, mathescape, breakatwhitespace=true, breaklines=true]
$\;\;\;\text{\textcolor{gray}{//Path: Dragonfly/src/}}\text{\textcolor{black}{\textbf{\underline{redis}}}}\text{\textcolor{gray}{/}}\text{\textcolor{black}{\textbf{\underline{lua}}}}\text{\textcolor{gray}{/struct/lua\_struct.c}}$
$\;\;\;\;\text{\textcolor{purple}{static int getnum (const char **fmt, int df) \{}}$
$\;\;\;\;\;\;\;\text{\textcolor{black}{...}}$
$\;\;\;\;\;\;\;\text{\textcolor{black}{int a = 0;}}$
$\;\;\;\;\;\;\;\text{\textcolor{black}{do \{}}$
$\;\;\;\;\;\;\;\;\;\;\;\text{\textcolor{purple}{a = a*10 + *((*fmt)++) - `0';}}$
\end{lstlisting}
\end{figure}

However, \texttt{Dragonfly} reuses \texttt{Redis}, and \texttt{Redis} does not use the upstream \texttt{Lua} as-is; it incorporates and modifies certain parts of the \texttt{Lua} codebase.
For example, CVE-2020-14147 was discovered by \texttt{Redis} in their modified \texttt{Lua} component. The vulnerability resides in \texttt{lua\_struct.c}, which does not exist in the upstream \texttt{Lua}. This integer overflow vulnerability was mitigated by validating the range of values assigned to the vulnerable variable \texttt{a} (\autoref{lst:redis}).

If the reuse genealogy 
\begin{tikzpicture}[
node/.style={draw, minimum width=0.3cm, minimum height=0.3cm, font=\small, inner sep=1pt},
arrow/.style={->, thick},
baseline = -0.7ex
]

\node[node] (lua) {\texttt{Lua}};
\node[node, right=0.5cm of lua] (redis) {\texttt{Redis}};
\node[node, right=0.5cm of redis] (fly) {\texttt{Dragonfly}};

\draw[arrow] (lua) -- (redis);
\draw[arrow] (redis) -- (fly);

\end{tikzpicture} is not identified,
existing SCA-based approaches miss the fact that \texttt{Dragonfly} inherited a vulnerability introduced by \texttt{Redis}, which is not present in the original \texttt{Lua}
(\autoref{lst:dragon}). This leads to blind spots in vulnerability detection and hampers timely patching in downstream projects.
Even fingerprint-based SCA tools (\eg, \cite{wu2023ossfp}) fail here, as \texttt{lua\_struct.c} does not exist in upstream \texttt{Lua} and thus no reference fingerprint exists for it.
Note that clone-based vulnerability detection approaches (\eg, \textsc{Movery}~\cite{woo2022movery}, FIRE~\cite{feng2024fire}) can detect this vulnerability, as the reused code remains largely unmodified.
However, \sys complements rather than replaces them: they identify point-in-time code similarity, whereas \sys reconstructs propagation paths across intermediate projects, reducing manual analysis overhead (see \autoref{subsec:app}).
This vulnerability remained unpatched until June 2025 and was subsequently fixed through our report.
\vspace{0.2em}
\section{Design of \sys}\label{sec:design}
\vspace{0.1em}

\subsection{Overview}\label{subsec:overview}

\begin{figure}[t]
	\begin{center}
		\includegraphics[width=0.95\linewidth]{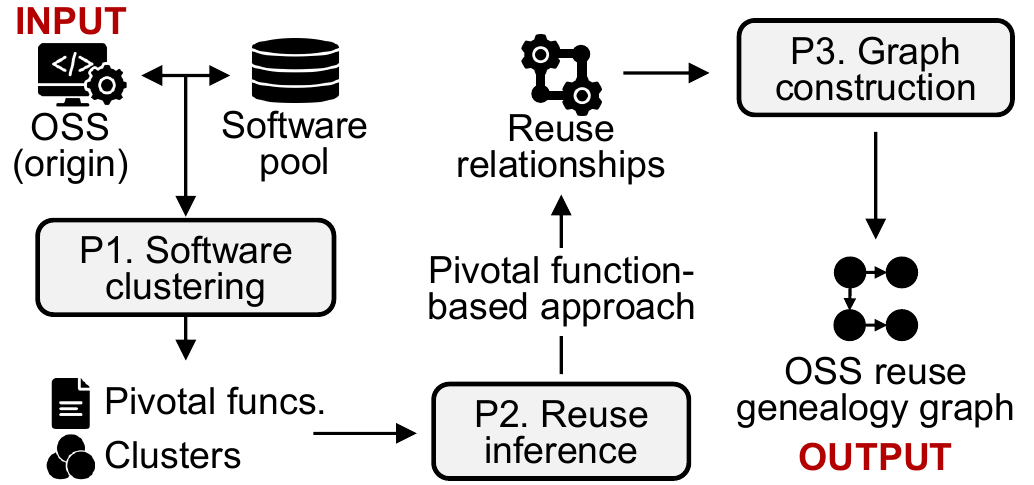}	
		
		\caption{\label{fig:overview}High-level overview of \sys.}
	\end{center}
	
\end{figure}

\autoref{fig:overview} illustrates the high-level workflow of \sys, which comprises three phases:
\textit{software clustering} (P1), \textit{reuse inference} (P2), and \textit{graph construction} (P3).
Given an OSS project (an origin), in P1, \sys identifies candidates within a pool of popular software projects and organizes them into clusters based on shared characteristics (\eg, reused file paths). In P2, \sys analyzes pairwise reuse relationships within each cluster using pivotal functions to measure similarity and determine reuse direction. In P3, \sys combines all relationships to construct a reuse genealogy graph, revealing propagation paths.

\subsection{Software Clustering (P1)}\label{subsec:p1}

\subsubsection{Software pool construction}\label{subsubsec:pool}
\sys first constructs a software pool from which the origin may have been reused, to identify its reuse genealogy.
To this end, \sys collects function codes and metadata
from popular software projects. \autoref{sec:eval} presents a detailed explanation of the software pool implementation.

\vspace{0.5em}
\begin{itemize}
    \setlength\itemsep{0.5em}
    \item \textbf{\textit{Function code.}}
    \sys collects the code of all functions belonging to each software. This is the primary basis for measuring code similarity between software projects.

    \item \textbf{\textit{Metadata of functions.}}
    \sys extracts two metadata elements for each function: (1) the release date of the version in which the function first appeared and (2) the associated file path.
    This metadata plays a crucial role in reuse analysis. 
\end{itemize}

\subsubsection{Candidate identification}\label{subsubsec:candi}
Based on the established software pool, \sys identifies candidate projects that may exhibit reuse relationships with the origin. To this end, \sys focuses on functions shared between the origin and each project in the pool.

To ensure robustness against changes that do not affect code semantics, \sys applies normalization to all functions contained in a software project, 
following practices from prior work (\eg, \cite{kim2017vuddy, xiao2020mvp, woo2022movery}). This includes removing whitespace, line breaks, and comments, as well as converting all characters to lowercase.

Note that \sys operates on raw source code prior to compiler preprocessing. Macro definitions and conditionally compiled blocks (\eg, \texttt{\#ifdef}) are treated as literal text during normalization and matching. Consequently, differences introduced at compile time, such as macro expansion or compiler flags, do not affect our comparison, which depends only on the normalized source text.

\sys then compares all normalized functions from the origin against those from each software function set in the software pool. To improve candidate detection accuracy, \sys extracts all functions from all versions of the origin that follow semantic versioning (\ie, \texttt{major.minor.patch}).

Subsequently,
\sys identifies syntactically \textit{identical} functions via exact string matching on normalized code, based on the principle that reuse relationships manifest through at least one identical function. Only projects sharing at least $\theta$ of the origin's functions are considered candidates, where $\theta$ is set to a low value (\eg, 1\%) to capture even cases where only a small portion of the origin (\eg, only several functions) is reused.
From these candidates, \sys derives the code regions on which the remaining analysis focuses.

\subsubsection{Pivotal files and functions}\label{subsubsec:pivotal}
We introduce two key concepts:

\vspace{0.1em}
\begin{itemize}
    \setlength\itemsep{0.4em}
        \item \textbf{Pivotal files.} We define a \textit{pivotal file} as a file containing at least one function identical to that of the origin.
        
        \item \textbf{Pivotal functions.} We define a \textit{pivotal function} as a function contained in a pivotal file.
\end{itemize}
\vspace{0.5em}

Existing SCA approaches compare entire codebases against the origin, which is not only inefficient but also introduces noise from unrelated code. \sys instead introduces pivotal functions, which confine the analysis to code regions derived from the origin; this scoping enables accurate pairwise comparison even among projects with large, heterogeneous codebases.

Let $X$ denote a software project, and $L_X$ represent the set of pivotal files within $X$.
We define $\mathcal{F}(l)$ as the set of functions contained in the file $l$. 
The set of all {\textit{pivotal functions} of $X$ ($F_X$) is given by:

\vspace{0.4em}
\begin{center}
$F_X = \bigcup\limits_{l \in L_X} \mathcal{F}(l)$
\end{center}
\vspace{0.4em}

Based on the identical functions identified in \autoref{subsubsec:candi}, \sys extracts pivotal files and functions for each candidate.

\subsubsection{Multi-criteria clustering}\label{subsubsec:clustering}

\sys then performs clustering based on the pivotal files and functions. This enables \sys to isolate unrelated candidates and focus reuse analysis on groups of related projects, thereby reducing false alarms.

For each pair of candidate projects $(X, Y)$, \sys examines the following four characteristics.

\begin{figure}[t]
	\begin{center}
		\includegraphics[width=1\linewidth]{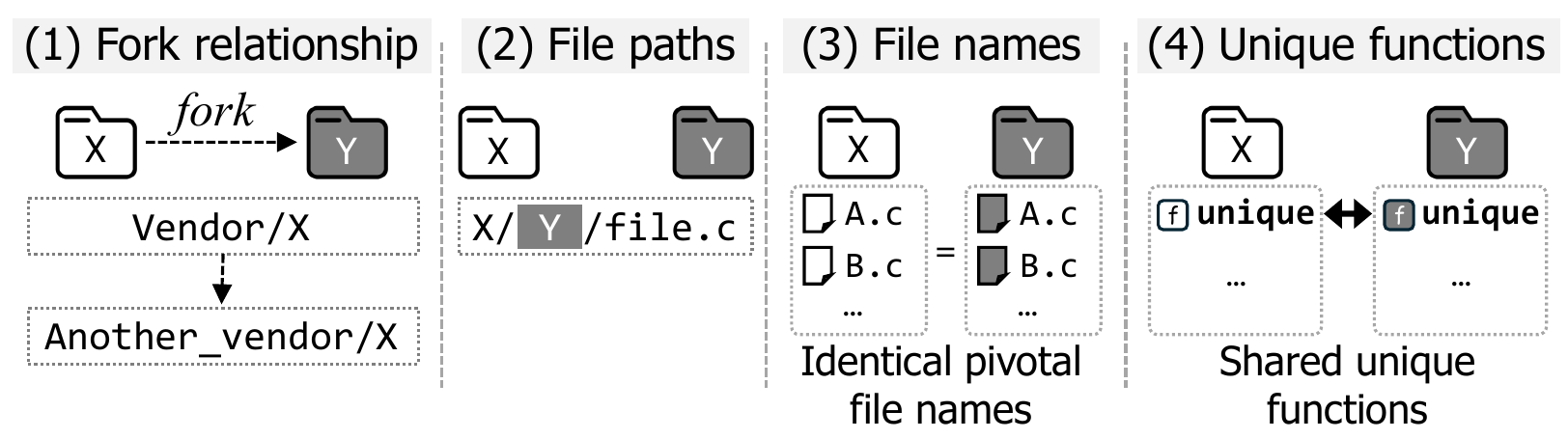}	
		
		\caption{\label{fig:cluster}Four criteria for clustering. If two candidates $X$ and $Y$ satisfy any criteria, they are grouped into the same cluster.}
	\end{center}

\end{figure}

\vspace{0.3em}
\begin{enumerate}
    \setlength\itemsep{0.4em}
    \item \textbf{\textit{Fork relationships.}}
    If either $X$ is a fork of $Y$ or \textit{vice versa}, they are grouped into the same cluster.

    \item \textbf{\textit{File paths.}}
    $X$ and $Y$ are clustered together if either project's name appears in any pivotal file path of the other.

    \item \textbf{\textit{File names.}}
    If the sets of pivotal file names between $X$ and $Y$ are identical, 
    then $X$ and $Y$ are placed in the same cluster.

    \item \textbf{\textit{Unique functions.}}
    A unique function is a pivotal function that does not exist in the origin. Projects sharing the same unique function are grouped into the same cluster.

\end{enumerate}
\vspace{0.3em}

\autoref{fig:cluster} illustrates these characteristics, which effectively cluster candidates with potential reuse relationships. 
Fork relationships and file paths are straightforward and reliable signals. Although file names alone may not indicate reuse, identical pivotal file names suggest that projects reused the same file set from the origin. Finally, a shared unique function suggests a reuse relationship, as it indicates that both projects inherited the same modification or addition to the origin.
If a pair satisfies any of the four criteria, the two projects are placed in the same cluster.

\subsection{Reuse Inference (P2)}\label{subsec:p2}
In P2, \sys infers reuse relationships among candidates within each cluster, in three steps: (1) measuring \textbf{code similarity} between two projects based on shared pivotal functions, (2) inferring the \textbf{reuse direction} using explicit and implicit information, and (3) aggregating the results to determine the overall \textbf{reuse relationship}.

\subsubsection{Code similarity measurement}
Let $X$ and $Y$ be two software projects within the same cluster. \sys computes not only the code similarity between $X$ and $Y$, but also between the origin $O$ and each of $X$ and $Y$. This allows \sys to differentiate three possible relationships: (1) $X$ is reused in $Y$ ($X \rightarrow Y$); (2) $Y$ is reused in $X$ ($Y \rightarrow X$); and (3) Both independently reuse $O$ ($O \rightarrow X$ and $O \rightarrow Y$).

In addition, \sys computes similarity by considering only pivotal functions
to focus on code regions derived from the origin.
\sys represents each project as a binary vector based on the presence or absence of pivotal functions, and computes similarity between the resulting embeddings.
To this end, \sys normalizes each pivotal function (see \autoref{subsubsec:candi}) and hashes it (\eg, \texttt{SHA256}).

Given two projects $X$ and $Y$, $F_X$ and $F_Y$ denote their respective sets of pivotal functions. Let $f_i$ represent a hashed function.
We define the union of these functions as the reference function set $F$:

\vspace{0.4em}
\begin{center}
$F = F_X \cup F_Y = \{f_1, f_2, \ldots, f_m\}$
\end{center}
\vspace{0.4em}

Each software is then embedded as a binary vector of dimension $m$ (let $\mathbf{v}_X$ and $\mathbf{v}_Y$), where the presence of function $f_i$ in project $X$ is encoded as follows.

\vspace{-0.4em}
\[
(\mathbf{v}_X)_i = \mathbb{I}(f_i \in F_X) =
\begin{cases}
1 & \text{if } f_i \in F_X, \\
0 & \text{otherwise}.
\end{cases}
\]
\vspace{0.2em}

\sys then measures the similarity  
using \textit{cosine similarity} ($\phi$). The cosine similarity between two vectors is calculated as follows.

\vspace{0.75em}
\begin{center} 
\small
 $\phi(\mathbf{v}_X, \mathbf{v}_Y) = \displaystyle\frac{\mathbf{v}_X \cdot \mathbf{v}_Y}{||\mathbf{v}_X|| \,||\mathbf{v}_Y||}$
\end{center}
\vspace{0.9em}

\sys focuses only on syntactically identical functions in $X$ and $Y$, based on the assumption that exact matches suffice for identifying reuse: in practice, when projects reuse code from a common origin, a substantial portion of functions remain unmodified~\cite{woo2021centris}, providing reliable evidence for inferring reuse direction.

Using this approach, \sys computes code similarity for every software pair 
($X$, $Y$) within a cluster, including the similarity between (1) $O$ and $X$, (2) $O$ and $Y$, and (3) $X$ and $Y$. 
An important consideration is that the origin $O$ does not have pivotal functions defined. To address this, \sys applies a reverse matching strategy: 
it identifies the functions in $O$ identical to the pivotal functions of $X$, and designates the corresponding files in $O$ as pivotal files. The similarity is then computed using the same methodology.

\begin{figure}[t]
	\begin{center}
		\includegraphics[width=1\linewidth]{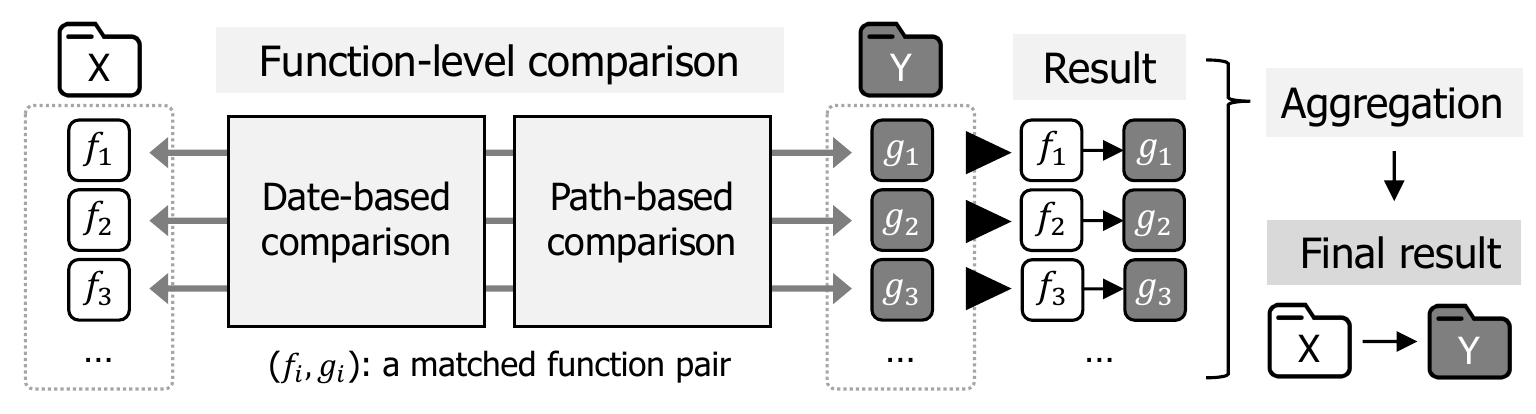}	
		
		\caption{\label{fig:p2process}Illustration of implicit matching based on function-level aggregation. 
        \sys captures fine-grained reuse patterns across functions and infers the reuse direction between software projects by aggregating this evidence.}
	\end{center}
\end{figure}

\subsubsection{Direction inference}
Next, \sys determines the direction of reuse between two projects $X$ and $Y$ in which a reuse relationship is \textit{assumed} to exist.
Note that determining reuse direction fundamentally distinguishes \sys from code clone detection: AST-based and token-based clone detectors identify a symmetric relation (whether two fragments are similar) but cannot determine which side is the ancestor. \sys resolves this asymmetry by leveraging \textit{explicit} and \textit{implicit} matches.

\PP{M1. Explicit matching}
First, \sys focuses on two explicit indicators that can directly confirm reuse relationships:
\textit{fork relationships} and \textit{file paths}.

\begin{itemize}
    \setlength\itemsep{0.35em}
    \item \textbf{Fork relationships.}
    If $Y$ is a fork of $X$, \sys infers a reuse direction from $X$ to $Y$. 
    
    \item \textbf{File paths.}
    If the path of a pivotal file in $Y$ contains the name of $X$, \sys infers that $Y$ reuses $X$. 
\end{itemize}

\PP{M2. Implicit matching}
If directionality cannot be identified through explicit matching, 
\sys applies implicit matching using function-level aggregation.
This method examines all matched pivotal function pairs ($f_i, g_i$) where $f_i \in F_X$ and $g_i \in F_Y$,
and aggregates evidence from individual functions to determine
the overall software-level reuse direction.

For each pair, 
\sys employs
two comparison methods.

\begin{itemize}
    \setlength\itemsep{0.4em}
    \item \textbf{Date-based comparison.} 
    \sys compares the release dates of the earliest versions in which $f_i$ and $g_i$ appeared in $X$ and $Y$, respectively. The function that appeared earlier is treated as the ancestor.
    
    \item \textbf{Path-based comparison.} \sys checks whether one function path fully contains the other. 
    In such cases, the function in the included path is regarded as the reused one, and the other as the ancestor.

\end{itemize}
\vspace{0.2em}

Here, version release dates can be inconsistent or even falsified, for example, when repository migrations or re-commits cause a downstream project's metadata to predate the origin. \sys addresses such unreliable timestamps through three mechanisms. First, explicit matching takes strict precedence over implicit matching; corrupted timestamps thus have no effect when explicit evidence exists. Second, path-based comparison provides a complementary signal even when date-based comparison is misleading. Third, \sys aggregates directional evidence across all matched function pairs via majority voting, so that a few corrupted timestamps are outvoted by consistent evidence from the remaining functions. Directionality inference fails only when explicit evidence, path signals, and consistent date evidence are all absent.

\autoref{fig:p2process} illustrates the function-level aggregation at a high level.
After each comparison, a directional decision is made (\eg, $f_i \rightarrow g_i$). If no clear direction can be determined (\eg, identical release dates), the pair is skipped. The overall reuse direction between projects is inferred based on majority voting across all matched function pairs:
\sys infers $X \rightarrow Y$ if more than half of the function pairs indicate $f_i \rightarrow g_i$.

If no direction is identified from \textit{any} of the functions,
\sys skips inferring a reuse relationship. Although this may suggest a fork relationship, the absence of explicit evidence from earlier matching prevents a confident decision.

\subsection{Graph Construction (P3)}\label{subsec:p3}
Finally, \sys constructs a reuse genealogy graph.

\vspace{0.7em}
\begin{center}
    \setlength{\fboxsep}{0.4em}
    \noindent\fbox{%
    \parbox{0.95\linewidth}{%
    	{{\scriptsize$\bullet$} \textbf{\ul{OSS reuse genealogy graph}}
        \vspace{4px}
        
        \hspace{6px}This is a directed acyclic graph that captures reuse relationships between software projects. An OSS reuse genealogy graph is defined as $G = (V, E)$, where $V$ is the set of nodes (software) and $E$ is the set of directed edges ($E \subseteq V \times V$). An edge $e = (v_1, v_2)$ indicates that $v_1$ is reused in $v_2$ ($v_1 \rightarrow v_2$).   
         }%
    }
    }
\end{center}
\vspace{0.7em}

Even with pairwise directions inferred, naively connecting all edges yields a noisy graph with redundant and conflicting paths, and no prior approach addresses how to consolidate pairwise relationships into a coherent genealogy. \sys resolves this with priority-based construction rules grounded in how code actually propagates in the OSS ecosystem (\eg, one-to-many reuse).

Specifically, the graph construction follows a principle of evidence strength and directness: explicit evidence takes precedence over implicit evidence, and among implicit connections, \sys prioritizes the strongest relationships to minimize noise.

\PP{Prioritizing explicit evidence.} Edges identified through explicit matching (fork or file paths) are always included regardless of similarity scores, as they provide definitive proof of reuse relationships.

\PP{Filtering implicit connections.} For relationships inferred through implicit matching, \sys applies systematic filtering. The priorities are assigned in descending order: R1, R2, R3, and R4.

\vspace{0.3em}
\begin{itemize}
    \setlength\itemsep{0.35em}
    \item[\textbf{R1.}] \textbf{Minimum similarity threshold.} Projects with low similarity scores ($\phi(X,Y) < \tau$, where $\tau$ is the threshold, \eg, 0.5) are not connected, as weak similarity is unreliable evidence of reuse.
    
    \item[\textbf{R2.}]
    \textbf{Strongest connection selection.} For each project, \sys connects only to the project with the highest similarity score. Because clear relationships are captured through explicit connections, focusing on the strongest implicit relationship avoids introducing noisy edges. When multiple candidates share the maximum score, all are considered.

    \item[\textbf{R3.}]
    \textbf{One-to-many propagation model.} \sys models realistic code propagation patterns: 
    one OSS project is frequently reused by multiple projects (one-to-many), but it is rare for developers to gather identical OSS code fragments from multiple projects and merge them into one (many-to-one).
    Therefore, when multiple edges point to the same project, only the strongest edge (\ie, highest similarity or explicit connection) is retained.

    \item[\textbf{R4.}]
    \textbf{Direct origin preference.}
    Finally, when $\phi(O, Y)$ is greater than $\phi(X, Y)$, this implies that $Y$ was more directly derived from $O$ rather than through intermediate $X$. Hence, \sys removes the $X\rightarrow Y$ edge to preserve the most direct genealogy path.
    
\end{itemize}

As an example, \autoref{table:graph_example} shows a subset of P2 results
for \texttt{zlib}.
Based on \textit{explicit} evidence, four edges are constructed.

\begin{table}[t]
\renewcommand{\tabcolsep}{0.5mm}	
	\caption{\label{table:graph_example}A partial result of similarity and direction inference within a cluster for \texttt{zlib} (the origin $O$).}
    \small

    \vspace{-0.95em}
\begin{center}

\begin{tabular}{|c|c|r|r|r|l|}
\hline
\rule{0in}{2.2ex}\textbf{$X$}
& \textbf{$Y$}  
& \multicolumn{1}{c|}{\textbf{$\phi(X, Y)$}}
& \multicolumn{1}{c|}{\textbf{$\phi(X, O)$}}
& \multicolumn{1}{c|}{\textbf{$\phi(O, Y)$}}
& \multicolumn{1}{c|}{\textbf{Direction}}\\\hline\hline
\rule{0in}{2.2ex}\texttt{FreeType}
& \texttt{Godot}
& 0.3543
& 0.3493
& 0.6643
& $X \rightarrow Y$ (Explicit)\\
\rule{0in}{2.2ex}\texttt{FreeType}
& \texttt{SumatraPDF}
& 0.1026
& 0.3493
& 0.5692
& $X \rightarrow Y$ (Explicit)\\
\rule{0in}{2.2ex}\texttt{FreeType}
& \texttt{Redot}
& 0.3112
& 0.3493
& 0.4329
& $X \rightarrow Y$ (Implicit)\\
\rule{0in}{2.2ex}\texttt{FreeType}
& \texttt{Miniblink49}
& 0.1071
& 0.3493
& 0.5036
& $X \rightarrow Y$ (Explicit)\\
\rule{0in}{2.2ex}\texttt{Godot}
& \texttt{Redot}
& 0.6282
& 0.6643
& 0.4329
& $X \rightarrow Y$ (Explicit)\\
\rule{0in}{2.2ex}\texttt{Godot}
& \texttt{SumatraPDF}
& 0.4986
& 0.6643
& 0.5692
& $X \rightarrow Y$ (Implicit)\\
\rule{0in}{2.2ex}\texttt{Godot}
& \texttt{Miniblink49}
& 0.5346
& 0.6643
& 0.5036
& $X \rightarrow Y$ (Implicit)\\
\rule{0in}{2.2ex}\texttt{SumatraPDF}
& \texttt{Redot}
& 0.1751
& 0.5692
& 0.4329
& $X \rightarrow Y$ (Implicit)\\
\rule{0in}{2.2ex}\texttt{SumatraPDF}
& \texttt{Miniblink49}
& 0.5674
& 0.5692
& 0.5036
& $Y \rightarrow X$ (Implicit)\\
\rule{0in}{2.2ex}\texttt{Redot}
& \texttt{Miniblink49}
& 0.1771
& 0.4329
& 0.5036
& $Y \rightarrow X$ (Implicit)\\\hline
\end{tabular}
\end{center}

\begin{flushleft}
\footnotesize{* All software versions are the latest as of June 2025.}
\end{flushleft}

\end{table}

\vspace{0.1em}
\begin{center}
    \small
    \setlength{\fboxsep}{0.4em}
    \noindent\fbox{%
    \parbox{0.95\linewidth}{%
    	{
        \textbf{$\boldsymbol{e_1}$.}
        \texttt{FreeType} $\rightarrow$ \texttt{Godot} (explicit)

        \vspace{0.1em} \textbf{$\boldsymbol{e_2}$.}
        \texttt{FreeType} $\rightarrow$ \texttt{SumatraPDF} (explicit)

        \vspace{0.1em}
        \textbf{$\boldsymbol{e_3}$.}
        \texttt{FreeType} $\rightarrow$ \texttt{Miniblink49} (explicit)

        \vspace{0.1em}
        \textbf{$\boldsymbol{e_4}$.}
        \texttt{Godot} $\rightarrow$ \texttt{Redot} (explicit)
        
        }
    } }
\end{center}
\vspace{0.2em}

Assume that $\tau = 0.5$. Excluding the edges created by explicit evidence, any edge with $\phi(X, Y) < \tau$ is disregarded according to \textbf{R1}. 
As a result, only the two edges remain among all implicit edges.

\vspace{0.45em}
{
\small
    \textbf{$\boldsymbol{e_5}$.}
    \texttt{Godot} $\rightarrow$ \texttt{Miniblink49} (edge candidate)

    \vspace{0.1em}
    
    \textbf{$\boldsymbol{e_6}$.}
    \texttt{Miniblink49} $\rightarrow$ \texttt{SumatraPDF} (edge candidate)     
}
\vspace{0.4em}

Next, among the two candidate edges ($e_5$ and $e_6$),
only $e_6$ satisfies \textbf{R2} and is retained, while $e_5$ is removed, as the similarity from \texttt{Godot} to \texttt{Redot} is higher than that from \texttt{Godot} to \texttt{Miniblink49}.
Also, 
because an explicit edge to \texttt{SumatraPDF} ($e_2$) already exists, $e_5$ should be removed (\textbf{R3}).
Although this case is not explicitly shown in the example, even if the similarity between \texttt{Godot} and \texttt{SumatraPDF} were 0.5, the edge would still not be established due to rule \textbf{R4}, as $\phi(O, Y)$ (0.5692) is higher than $\phi(X, Y)$ (0.5).
Consequently, \textbf{only the edges $\boldsymbol{e_1}$ through $\boldsymbol{e_4}$ remain}.

\begin{figure}[t]
	\begin{center}
		\includegraphics[width=0.8\linewidth]{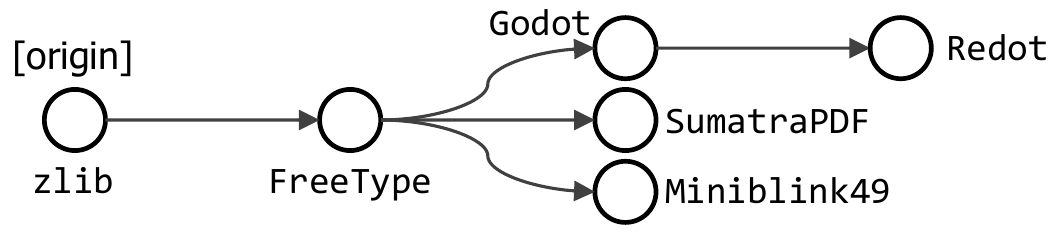}	
		
		\vspace{-0.5em}
		\caption{\label{fig:working_graph}OSS reuse genealogy graph generated from \autoref{table:graph_example}.}
	\end{center}
	
	\vspace{-0.7em}
\end{figure}

Through this process, the reuse genealogy among the projects within each cluster is determined. 
As a final step, \sys connects the origin to all nodes with zero indegree, ensuring that every project can be traced back to the origin. This includes both software projects with no direct reuse relationships and starting points of existing reuse chains.
Individual software projects that do not belong to any cluster are also connected to the origin, allowing \sys to produce a comprehensive reuse genealogy.

\autoref{fig:working_graph} shows the graph generated based on \autoref{table:graph_example}.
\sys successfully reconstructs the reuse genealogy of \texttt{zlib}: \texttt{FreeType} reuses code from \texttt{zlib}; \texttt{Godot}, \texttt{SumatraPDF}, and \texttt{Miniblink49} inherit \texttt{zlib} component code through \texttt{FreeType}; and 
\texttt{Redot} inherits \texttt{zlib} from \texttt{Godot}, as it was developed as a fork of \texttt{Godot}.
\section{Evaluation}\label{sec:eval}

In this section, we experimentally evaluate \sys to answer the
following three research questions.

\vspace{0.3em}
\begin{enumerate}
    \setlength\itemsep{0.4em}
    \item [\textbf{RQ1.}] 
    \textbf{Accuracy.}  
    How precisely and effectively does \sys identify OSS reuse genealogy? (\autoref{subsec:accuracy})

    \item [\textbf{RQ2.}] \textbf{Performance and scalability.}
    How efficient and scalable is \sys?
    (\autoref{subsec:performance})

    \item [\textbf{RQ3.}] \textbf{Application.}
    How does \sys contribute to securing the software supply chain?
    (\autoref{subsec:app}) 

\end{enumerate}
\vspace{0.32em}

We ran \sys on an AWS m7i.4xlarge instance running Ubuntu 22.04.5 LTS, equipped with an Intel Xeon Platinum 8488C Processor (16 vCPUs, 2.4GHz), 64GB RAM, and a 2TB NVMe SSD.

\PP{Implementation of \sys}
\sys comprises approximately 1,300 lines of Python code,  excluding external libraries such as \texttt{Tree-sitter}
and \texttt{scikit-learn}~\cite{pedregosa2011scikit}. It includes three modules: a \textit{dataset collector} that builds the software pool, a \textit{clusterer} that identifies and groups candidates (see \autoref{subsec:p1}), and a \textit{reuse genealogy identifier} that infers reuse directions and constructs the genealogy graph (\autoref{subsec:p2} and \autoref{subsec:p3}).

\PP{Software pool construction}
To build the software pool, 
\sys collected all versions of the top 2,500 C/C++ repositories on GitHub ranked by stargazers (as of June 2025),
including prominent projects such as \texttt{Linux} \texttt{Kernel}, \texttt{Wireshark}, and \texttt{Redis}. 
\sys then extracted all functions from each repository using \texttt{Tree-sitter}.
Following prior work (\eg, \cite{woo2021centris,choi2025tiver}), we consider versions based on \texttt{GitHub} tags, but only include those that follow \texttt{major.minor.patch} semantic versioning.
Release dates were retrieved using 
\texttt{Git} commands (\eg, \texttt{git} \texttt{tag} \texttt{-{}-sort=creatordate}). 
Repositories without tags or non-semantic versioning were excluded due to the lack of reliable release date information.
Consequently, \sys stored codebases and metadata for 2,006 repositories in the pool.

\subsection{Accuracy of \sys}\label{subsec:accuracy}

We first evaluate the genealogy construction accuracy of \sys on widely reused OSS projects. 
\subsubsection{Target software selection}
To avoid bias in target OSS selection, we consider the following three criteria:
(1) the OSS should be popular and thus well-maintained,
(2) the selected OSS projects should collectively provide at least 1,000 propagation paths within the pool for reliable accuracy evaluation, and
(3) the target OSS projects should span diverse domains.

To this end, we referred to the results of \cent~\cite{woo2021centris}, which analyzed OSS components for 15,000 popular projects. We filtered these results to identify prime OSS (\ie, projects that do not incorporate other components) and sorted them by reuse frequency. Here, the top 20 OSS projects collectively provided over 1,000 reuse paths, satisfying the second criterion; we thus selected them as evaluation targets (see \autoref{table:acc}).
This set includes popular OSS such as \texttt{Lua}, \texttt{Libxml2}, and \texttt{json-c}, and spans diverse domains including compression, file systems, and media processing.

\vspace{0.1em}
\subsubsection{Accuracy comparison with \cn}
We first measure and compare the accuracy of reuse genealogy identification between \sys and \cn~\cite{na2024cneps}, which analyzes component dependencies. 

\vspace{5px}
\noindent\textbf{Adaptation of \cn for evaluation.}
Although \cn does not directly identify OSS reuse genealogies, it infers the dependency direction among reused components.
For example, given \texttt{Godot} as input (see \autoref{fig:working_graph}), \cn identifies that \texttt{FreeType} and \texttt{zlib} are reused, yielding
\begin{tikzpicture}[
node/.style={draw, minimum width=0.3cm, minimum height=0.3cm, font=\small, inner sep=1pt},
arrow/.style={->, thick},
baseline = -0.7ex
]

\node[node] (godot) {\texttt{Godot}};
\node[node, right=0.5cm of godot] (freetype) {\texttt{Freetype}};
\node[node, right=0.5cm of freetype] (zlib) {\texttt{zlib}};

\draw[arrow] (godot) -- (freetype);
\draw[arrow] (freetype) -- (zlib);

\end{tikzpicture}.

Therefore, we construct a reverse dependency graph from the \cn results. We invert the direction of each dependency edge so that the component that originally appears at the end of the dependency chain becomes the starting point of the graph, as follows:
\begin{tikzpicture}[
node/.style={draw, minimum width=0.3cm, minimum height=0.3cm, font=\small, inner sep=1pt},
arrow/.style={->, thick},
baseline = -0.7ex
]

\node[node] (godot) {\texttt{Godot}};
\node[node, right=0.5cm of godot] (freetype) {\texttt{Freetype}};
\node[node, right=0.5cm of freetype] (zlib) {\texttt{zlib}};

\draw[arrow] (freetype) -- (godot);
\draw[arrow] (zlib) -- (freetype);

\end{tikzpicture} (reverse \cn results).
This yields a structure that approximates the reuse propagation relationships.

We initially executed \cn on the 2,006 projects in our pool. However, when queried with a downstream project $X$, \cn sometimes failed to reveal that $X$ reuses the target OSS. In such cases, the relationship might be recoverable indirectly: querying a project $Y$ (a downstream of $X$ outside our pool) returned the chain OSS$\rightarrow$$X$$\rightarrow$$Y$, from which the missing relationship could be extracted. We therefore executed \cn on approximately 10,000 projects from the \cn dataset to maximize coverage, and retained only the paths within our pool (\eg, discarding the $X$$\rightarrow$$Y$ edge for $Y$ outside the pool), ensuring a fair comparison.
For each project, we used the version released as of June 2025, as analyzing multiple versions increases computational cost while yielding similar results.
\vspace{0.3em}

\PP{Methodology}
We assessed the accuracy of \sys and \cn by analyzing the reuse directions in the genealogy graphs. 
However, validation remains challenging due to the absence of benchmark datasets for C/C++, where reuse relationships are rarely documented.
Therefore, we manually analyzed all the results. The evaluation is conducted by two experts: one with over 15 years and the other with over five years of experience in software engineering and security.
We primarily rely on pivotal file paths or metadata (\eg, \texttt{README}) that provide clues about the origin.
We then analyze source code and comments, as well as the commit history, to assess the accuracy of reuse relationships as thoroughly as possible.
The decision is reached through discussion between the two evaluators.

We consider five evaluation metrics: \textit{true positives} (\texttt{TP}), \textit{false positives} (\texttt{FP}), 
\textit{false negatives} (\texttt{FN}),
\textit{precision} (\#\texttt{TP}/(\#\texttt{TP} + \#\texttt{FP})), and \textit{recall} (\#\texttt{TP}/(\#\texttt{TP} + \#\texttt{FN})).
A TP is a correctly identified reuse relationship, an FP is an incorrect inference, and an FN is a missed reuse relationship.
\textit{True negatives} are excluded, because the number of non-reused projects is overwhelmingly large, potentially introducing bias.
We set the threshold $\theta$ (see \autoref{subsubsec:candi}) for candidate identification to the minimum value (\ie, projects are considered candidates if they share at least one function with the target OSS), and set the threshold $\tau$ (see \autoref{subsec:p3}) for reuse relationship analysis to 0.5. Experiments on threshold sensitivity are presented in
\autoref{subsubsec:threshold}.

\begin{table}[t]
	\centering
	\small
	\renewcommand{\tabcolsep}{0.47mm}
	\caption{\label{table:acc}Accuracy measurement results of \cn and \sys on 20 widely reused OSS projects (P: precision, R: recall).}
	\vspace{-1em}
    \begin{tabular}{|c||rrrrr||rrrrr|}
    \hline
    \multirow{2}{*}{\begin{tabular}[c|]{@{}c@{}}\rule{0in}{2.2ex}\textbf{Target}\\\textbf{OSS}\end{tabular}} 
    & \multicolumn{5}{c||}{\rule{0in}{2ex}\textbf{\cn~\cite{na2024cneps}}}   
    & \multicolumn{5}{c|}{\rule{0in}{2ex}\textbf{\sys}}\\\cline{2-11} 
    \rule{0in}{2ex}
    & \multicolumn{1}{c|}{\rule{0in}{2.2ex}\textbf{\#TP}} 
    & \multicolumn{1}{c|}{\textbf{\#FP}}
    & \multicolumn{1}{c|}{\textbf{\#FN}} 
    & \multicolumn{1}{c|}{\textbf{P (\%)}}
    & \multicolumn{1}{c||}{\textbf{R (\%)}} 
    & \multicolumn{1}{c|}{\textbf{\#TP}} 
    & \multicolumn{1}{c|}{\textbf{\#FP}}
    & \multicolumn{1}{c|}{\textbf{\#FN}} 
    & \multicolumn{1}{c|}{\textbf{P (\%)}} 
    & \multicolumn{1}{c|}{\textbf{R (\%)}} \\\hline\hline
    \texttt{json-c} 
    & \multicolumn{1}{c|}{2}  
    & \multicolumn{1}{c|}{0}  
    & \multicolumn{1}{c|}{607}  
    & \multicolumn{1}{r|}{100.00}
    & \multicolumn{1}{r||}{0.33}                     
    & \multicolumn{1}{c|}{594}  
    & \multicolumn{1}{c|}{49}  
    & \multicolumn{1}{c|}{15}  
    & \multicolumn{1}{r|}{92.38}  
    & 97.54\\
    \texttt{stb} 
    & \multicolumn{1}{c|}{21}  
    & \multicolumn{1}{c|}{2}  
    & \multicolumn{1}{c|}{146}  
    & \multicolumn{1}{r|}{91.30}
    & \multicolumn{1}{r||}{12.57}                     
    & \multicolumn{1}{c|}{160}  
    & \multicolumn{1}{c|}{64}  
    & \multicolumn{1}{c|}{7}  
    & \multicolumn{1}{r|}{71.43}  
    & 95.81\\
    \texttt{zlib} 
    & \multicolumn{1}{c|}{146}  
    & \multicolumn{1}{c|}{12}  
    & \multicolumn{1}{c|}{15}  
    & \multicolumn{1}{r|}{92.41}
    & \multicolumn{1}{r||}{90.68}                     
    & \multicolumn{1}{c|}{148}  
    & \multicolumn{1}{c|}{39}  
    & \multicolumn{1}{c|}{13}  
    & \multicolumn{1}{r|}{79.14}  
    & 91.93\\
    \texttt{libsodium} 
    & \multicolumn{1}{c|}{3}  
    & \multicolumn{1}{c|}{0}  
    & \multicolumn{1}{c|}{123}  
    & \multicolumn{1}{r|}{100.00}
    & \multicolumn{1}{r||}{2.38}                     
    & \multicolumn{1}{c|}{123}  
    & \multicolumn{1}{c|}{24}  
    & \multicolumn{1}{c|}{3}  
    & \multicolumn{1}{r|}{83.67}  
    & 97.62\\
    \texttt{libuv} 
    & \multicolumn{1}{c|}{7}  
    & \multicolumn{1}{c|}{1}  
    & \multicolumn{1}{c|}{89}  
    & \multicolumn{1}{r|}{87.50}
    & \multicolumn{1}{r||}{7.29}                     
    & \multicolumn{1}{c|}{94}  
    & \multicolumn{1}{c|}{9}  
    & \multicolumn{1}{c|}{2}  
    & \multicolumn{1}{r|}{91.26}  
    & 97.92\\
    \texttt{xxHash} 
    & \multicolumn{1}{c|}{69}  
    & \multicolumn{1}{c|}{6}  
    & \multicolumn{1}{c|}{9}  
    & \multicolumn{1}{r|}{92.00}
    & \multicolumn{1}{r||}{88.46}                     
    & \multicolumn{1}{c|}{75}  
    & \multicolumn{1}{c|}{16}  
    & \multicolumn{1}{c|}{3}  
    & \multicolumn{1}{r|}{82.42} 
    & 96.15\\ 
    \texttt{Lua} 
    & \multicolumn{1}{c|}{60}  
    & \multicolumn{1}{c|}{7}  
    & \multicolumn{1}{c|}{17}  
    & \multicolumn{1}{r|}{89.55}
    & \multicolumn{1}{r||}{77.92}                     
    & \multicolumn{1}{c|}{68}  
    & \multicolumn{1}{c|}{25}  
    & \multicolumn{1}{c|}{9}  
    & \multicolumn{1}{r|}{73.12} 
    & 88.31\\ 
    \texttt{Expat} 
    & \multicolumn{1}{c|}{20}  
    & \multicolumn{1}{c|}{2}  
    & \multicolumn{1}{c|}{17}  
    & \multicolumn{1}{r|}{90.91}
    & \multicolumn{1}{r||}{54.05}                   
    & \multicolumn{1}{c|}{35}  
    & \multicolumn{1}{c|}{5}  
    & \multicolumn{1}{c|}{2}  
    & \multicolumn{1}{r|}{87.50} 
    & 94.59\\ 
    \texttt{flex} 
    & \multicolumn{1}{c|}{4}  
    & \multicolumn{1}{c|}{1}  
    & \multicolumn{1}{c|}{28}  
    & \multicolumn{1}{r|}{80.00}
    & \multicolumn{1}{r||}{12.50}                   
    & \multicolumn{1}{c|}{29}  
    & \multicolumn{1}{c|}{8}  
    & \multicolumn{1}{c|}{3}  
    & \multicolumn{1}{r|}{78.38} 
    & 90.63\\ 
    \texttt{PCRE2} 
    & \multicolumn{1}{c|}{0}  
    & \multicolumn{1}{c|}{0}  
    & \multicolumn{1}{c|}{29}  
    & \multicolumn{1}{r|}{0.00}
    & \multicolumn{1}{r||}{0.00}                   
    & \multicolumn{1}{c|}{27}  
    & \multicolumn{1}{c|}{6}  
    & \multicolumn{1}{c|}{2}  
    & \multicolumn{1}{r|}{81.82} 
    & 93.10\\ 
    \texttt{libunwind} 
    & \multicolumn{1}{c|}{3}  
    & \multicolumn{1}{c|}{0}  
    & \multicolumn{1}{c|}{24}  
    & \multicolumn{1}{r|}{100.00}
    & \multicolumn{1}{r||}{11.11}                   
    & \multicolumn{1}{c|}{26}  
    & \multicolumn{1}{c|}{1}  
    & \multicolumn{1}{c|}{1}  
    & \multicolumn{1}{r|}{96.30} 
    & 96.30\\ 
    \texttt{Ogg} 
    & \multicolumn{1}{c|}{0}  
    & \multicolumn{1}{c|}{0}  
    & \multicolumn{1}{c|}{24}  
    & \multicolumn{1}{r|}{0.00}
    & \multicolumn{1}{r||}{0.00}                   
    & \multicolumn{1}{c|}{23}  
    & \multicolumn{1}{c|}{3}  
    & \multicolumn{1}{c|}{1}  
    & \multicolumn{1}{r|}{88.46} 
    & 95.83\\ 
    \texttt{Libxml2} 
    & \multicolumn{1}{c|}{5}  
    & \multicolumn{1}{c|}{0}  
    & \multicolumn{1}{c|}{16}  
    & \multicolumn{1}{r|}{100.00}
    & \multicolumn{1}{r||}{23.81}                   
    & \multicolumn{1}{c|}{20}  
    & \multicolumn{1}{c|}{3}  
    & \multicolumn{1}{c|}{1}  
    & \multicolumn{1}{r|}{86.96} 
    & 95.24\\ 
    \texttt{Vorbis} 
    & \multicolumn{1}{c|}{0}  
    & \multicolumn{1}{c|}{0}  
    & \multicolumn{1}{c|}{22}  
    & \multicolumn{1}{r|}{0.00}
    & \multicolumn{1}{r||}{0.00}                   
    & \multicolumn{1}{c|}{19}  
    & \multicolumn{1}{c|}{11}  
    & \multicolumn{1}{c|}{3}  
    & \multicolumn{1}{r|}{63.33} 
    & 86.36\\ 
    \texttt{TZ} 
    & \multicolumn{1}{c|}{15}  
    & \multicolumn{1}{c|}{1}  
    & \multicolumn{1}{c|}{6}  
    & \multicolumn{1}{r|}{93.75}
    & \multicolumn{1}{r||}{71.43}                   
    & \multicolumn{1}{c|}{19}  
    & \multicolumn{1}{c|}{2}  
    & \multicolumn{1}{c|}{2}  
    & \multicolumn{1}{r|}{90.48} 
    & 90.48\\ 
    \texttt{libzip} 
    & \multicolumn{1}{c|}{3}  
    & \multicolumn{1}{c|}{0}  
    & \multicolumn{1}{c|}{16}  
    & \multicolumn{1}{r|}{100.00}
    & \multicolumn{1}{r||}{15.79}                   
    & \multicolumn{1}{c|}{19}  
    & \multicolumn{1}{c|}{1}  
    & \multicolumn{1}{c|}{0}  
    & \multicolumn{1}{r|}{95.00} 
    & 100.00\\ 
    \texttt{libpcap} 
    & \multicolumn{1}{c|}{2}  
    & \multicolumn{1}{c|}{2}  
    & \multicolumn{1}{c|}{15}  
    & \multicolumn{1}{r|}{50.00}
    & \multicolumn{1}{r||}{11.76}                   
    & \multicolumn{1}{c|}{16}  
    & \multicolumn{1}{c|}{6}  
    & \multicolumn{1}{c|}{1}  
    & \multicolumn{1}{r|}{72.73} 
    & 94.12\\ 
    \texttt{libCoAP} 
    & \multicolumn{1}{c|}{1}  
    & \multicolumn{1}{c|}{0}  
    & \multicolumn{1}{c|}{14}  
    & \multicolumn{1}{r|}{100.00}
    & \multicolumn{1}{r||}{6.67}                   
    & \multicolumn{1}{c|}{15}  
    & \multicolumn{1}{c|}{2}  
    & \multicolumn{1}{c|}{0}  
    & \multicolumn{1}{r|}{88.24} 
    & 100.00\\ 
    \texttt{LuaBitOP} 
    & \multicolumn{1}{c|}{11}  
    & \multicolumn{1}{c|}{4}  
    & \multicolumn{1}{c|}{3}  
    & \multicolumn{1}{r|}{73.33}
    & \multicolumn{1}{r||}{78.57}                   
    & \multicolumn{1}{c|}{14}  
    & \multicolumn{1}{c|}{0}  
    & \multicolumn{1}{c|}{0}  
    & \multicolumn{1}{r|}{100.00} 
    & 100.00\\ 
    \texttt{libui} 
    & \multicolumn{1}{c|}{0}  
    & \multicolumn{1}{c|}{0}  
    & \multicolumn{1}{c|}{11}  
    & \multicolumn{1}{r|}{0.00}
    & \multicolumn{1}{r||}{0.00}                   
    & \multicolumn{1}{c|}{11}  
    & \multicolumn{1}{c|}{0}  
    & \multicolumn{1}{c|}{0}  
    & \multicolumn{1}{r|}{100.00} 
    & 100.00\\\hline\hline
    \texttt{\textbf{Total}} 
    & \multicolumn{1}{c|}{372}  
    & \multicolumn{1}{c|}{\textcolor{tomato}{\textbf{38}}}  
    & \multicolumn{1}{c|}{1,231}  
    & \multicolumn{1}{r|}{\textcolor{tomato}{\textbf{90.73}}}
    & \multicolumn{1}{r||}{23.21}                   
    & \multicolumn{1}{c|}{\textcolor{tomato}{\textbf{1,535}}}  
    & \multicolumn{1}{c|}{274}  
    & \multicolumn{1}{c|}{\textcolor{tomato}{\textbf{68}}}  
    & \multicolumn{1}{r|}{84.85} 
    & \textcolor{tomato}{\textbf{95.76}}\\\hline
\end{tabular}
\end{table}

\begin{table}[t]
\centering
\renewcommand{\tabcolsep}{3.8mm}	
    \caption{\label{table:cn} 
    Distribution of propagation depths of reuse relationships identified by \cn and \sys.}
    \small
    \vspace{-1em}
\begin{tabular}{|c||c|c|c|c|c||c|}
\hline
\multirow{2}{*}{\rule{0in}{2.2ex}\textbf{Tool}}
& \multicolumn{5}{c||}{\rule{0in}{2.2ex}\textbf{Propagation depth}} 
& \multirow{2}{*}{\rule{0in}{2.2ex}\textbf{Total}}\\\cline{2-6}
& \multicolumn{1}{c|}{\rule{0in}{2.2ex}\textbf{2}} 
& \multicolumn{1}{c|}{\textbf{3}} 
& \multicolumn{1}{c|}{\textbf{4}} 
& \multicolumn{1}{c|}{\textbf{5}} 
& \multicolumn{1}{c||}{\textbf{6}}                 
& \\\hline\hline
\rule{0in}{2ex}\cn 
& 298
& 68                   
& 5                    
& 1                     
& 0                     
& 372 \\\hline
\rule{0in}{2ex}\sys 
& 1,169
& 354                 
& 8                   
& 3                     
& 1                     
& 1,535\\\hline   
\end{tabular}

\vspace{-0.4em}
\end{table}

\vspace{0.3em}
\PP{Overall results}
\autoref{table:acc} shows the results.
More than 1,600 correct reuse relationships were identified; over 20\% involved propagation through at least one intermediate project, 
with the longest chain passing through four \textit{intermediate} projects (\autoref{table:cn}).
Despite these challenging transitive cases, \sys achieved 84.85\% precision and 95.76\% recall, whereas \cn achieved higher precision (90.73\%) but considerably lower recall (23.21\%).

\vspace{7px}
\noindent\textbf{Result analysis: \cn.}
Although \cn generated few FPs, it yielded many FNs (23.21\% recall).
\cn reported fewer FPs because it identifies OSS components based on \cent and traces clone paths within files to filter out incidental code similarity, thereby considering only clear and verifiable reuse cases.
However, this design also causes the FNs: \cn focuses on one-depth dependencies rather than transitive paths, and emphasizes only explicit reuse patterns (\eg, \texttt{\#include}), overlooking many indirect or implicit reuse relationships.
In particular, \cn failed to identify reuse of \texttt{json-c}, \texttt{stb}, and \texttt{libsodium} as components.

\vspace{3.5px}
\noindent\textbf{Result analysis: \sys.} 
\sys successfully identified even transitive OSS reuse.
For example, in the case of \texttt{stb}, we found that \texttt{Filament} reuses \texttt{stb} both directly and indirectly through \texttt{glfw} (\texttt{stb} $\rightarrow$ \texttt{glfw} $\rightarrow$ \texttt{Filament} and \texttt{stb} $\rightarrow$ \texttt{Filament}).
In particular, \texttt{Filament} includes \texttt{stb} code in its own source tree
and through \texttt{glfw}
(\path{/third_party/glfw/deps/stb_image_write.h}).

However, some FPs and FNs were observed. FPs arose from three causes: (1) missing intermediate projects, (2) reuse of identical origin versions with similar path structures, and (3) commonly shared code snippets (\eg, cryptographic code). For the first case, if $Y$ is missing from the pool in a chain $X\rightarrow Y\rightarrow Z$, an incorrect relationship such as $X\rightarrow Z$ may be inferred.
The latter two cases occur when unrelated projects reused the same origin version with similar path structures, or when only short, generic code was shared across projects.
Most FNs occurred when the reuse relationship was implicit and the proportion of shared functions fell below $\tau$.

\begin{table}[t]
\renewcommand{\tabcolsep}{0.85mm}	
    \caption{\label{table:v0}Accuracy comparison (\sys \textit{vs.} \vzfinder).}
    \small

    \vspace{-1em}
\begin{center}

\begin{tabular}{|c|c|c|c|r|r|c|c|c|r|r|}
\hline
\multirow{2}{*}{\begin{tabular}[c]{@{}c@{}}\rule{0in}{2.2ex}\textbf{Target}\\\textbf{OSS}\end{tabular}}
& \multicolumn{5}{c|}{\rule{0in}{2.2ex}\textbf{\vzfinder}~\cite{woo2021v0finder}} 
& \multicolumn{5}{c|}{\rule{0in}{2.2ex}\textbf{\sys}}\\\cline{2-11}
& \multicolumn{1}{c|}{\rule{0in}{2.2ex}\textbf{\#TP}}
& \multicolumn{1}{c|}{\rule{0in}{2.2ex}\textbf{\#FP}}
& \multicolumn{1}{c|}{\rule{0in}{2.2ex}\textbf{\#FN}}
& \multicolumn{1}{c|}{\rule{0in}{2.2ex}\textbf{P(\%)}}
& \multicolumn{1}{c|}{\rule{0in}{2.2ex}\textbf{R(\%)}}
& \multicolumn{1}{c|}{\rule{0in}{2.2ex}\textbf{\#TP}}
& \multicolumn{1}{c|}{\rule{0in}{2.2ex}\textbf{\#FP}}
& \multicolumn{1}{c|}{\rule{0in}{2.2ex}\textbf{\#FN}}
& \multicolumn{1}{c|}{\rule{0in}{2.2ex}\textbf{P(\%)}}
& \multicolumn{1}{c|}{\rule{0in}{2.2ex}\textbf{R(\%)}}

\\\hline\hline
\texttt{Lua}
&4
&7
&73
&36.36
&5.19
&68
&25
&9
&73.12
&88.31\\
\texttt{Expat}
&17
&20
&20
&45.95
&45.95
&35
&5
&2
&87.50
&94.59\\
\texttt{Ogg}
&4
&0
&20
&100.00
&16.67
&23
&3
&1
&88.46
&95.83\\
\texttt{Libxml2}
&4
&3
&17
&57.14
&19.05
&20
&3
&1
&86.96
&95.24\\
\texttt{Vorbis}
&10
&0
&12
&100.00
&45.45
&19
&11
&3
&63.33
&86.36\\\hline\hline
\rule{0in}{2.2ex}\textbf{Total}
&39
&30
&142
&56.52
&21.55
&165
&47
&16
&\textcolor{tomato}{\textbf{77.83}}
&\textcolor{tomato}{\textbf{91.16}}\\\hline

\end{tabular}
\end{center}

\end{table}

\begin{figure}[t]
	
	\begin{center}
		\begin{subfigure}[b]{0.14\textwidth}
			\centering
			\includegraphics[width=\linewidth]{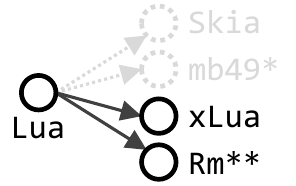}
	
			\vspace{-0.3em}
			\caption{\vzfinder}
			\label{fig:eff_v0}
		\end{subfigure}\hspace{1em}%
		\begin{subfigure}[b]{0.14\textwidth}
			\centering
			\includegraphics[width=\linewidth]{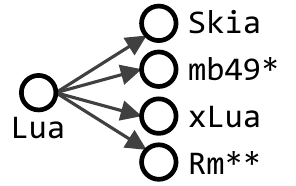}
			
			\vspace{-0.3em}
			\caption{\cn}
			\label{fig:eff_cn}
		\end{subfigure}\hspace{1em}%
		\begin{subfigure}[b]{0.14\textwidth}
			\centering
			\includegraphics[width=\linewidth]{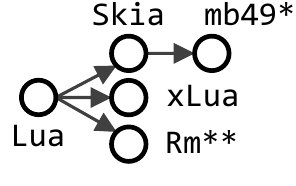}
			
			\vspace{-0.3em}
			\caption{\sys}
			\label{fig:eff_ganadi}
		\end{subfigure}
		\vspace{-0.6em}
  
		\caption{Key differences among \vzfinder, \cn, and \sys (\texttt{mb49}: \texttt{miniblink49}, \texttt{Rm}: \texttt{Rainmeter}).}\label{fig:eff}
	\end{center}
	
	\vspace{-1em}
\end{figure}

\subsubsection{Accuracy comparison with \vzfinder} 
We next compare the accuracy of \sys with that of \vzfinder~\cite{woo2021v0finder}.
Although identifying reuse relationships is not the primary objective of \vzfinder, it relies on reuse information to detect vulnerability propagation.

\PP{Methodology} 
Unlike \cn, \vzfinder is limited to projects with known vulnerabilities. We thus restricted our evaluation to projects that satisfy its requirements:
(1) a reported CVE exists, and (2) the corresponding patch is available as a \texttt{GitHub} commit. 
As a result, five OSS projects were selected.

\PP{Result analysis}
\autoref{table:v0} presents the accuracy comparison results.
The most significant difference lies in recall: \sys achieved 91.16\%, while \vzfinder reached only 21.55\%.
This gap is caused by two main limitations of \vzfinder: 
it (1) detects reuse only when vulnerable code is propagated and (2) infers reuse based on coarse signals (\eg, overall code similarity).
Although \vzfinder produced fewer FPs (30 \textit{vs.} \sys's 47), this is primarily due to its overly strict criteria for identifying reuse edges.
Supporting this, it identified only 39 TPs, significantly fewer than \sys (165 TPs).

Overall, \cn identifies origins but lacks accuracy for detailed reuse relationships, while \vzfinder is limited to cases where vulnerabilities have propagated.  \autoref{fig:eff} illustrates these differences.

\subsubsection{Ablation study}
To quantify the contribution of each component, we removed one component at a time across the three phases and measured the accuracy on the 20 target OSS projects.
Specifically, when a clustering criterion is removed, candidate pairs are no longer grouped by that criterion; when explicit (or implicit) direction matching is removed, directions are inferred using only the remaining method; and when explicit (or implicit) edges are removed, the graph is constructed using only the other type.

\autoref{table:ablation} presents the results.
Removing any single component degraded both precision and recall, indicating that every component is necessary.
The largest degradation occurred when explicit edges were removed from graph construction (58.39\% precision and 52.96\% recall): relationships evidenced by forks or file paths often exhibit low code similarity, and without explicit prioritization, they were discarded by the similarity threshold $\tau$.
Conversely, removing implicit edges lowered recall to 67.19\%, as only the relationships confirmed by explicit evidence survived; it also introduced FPs, because when an intermediate edge is lost (\eg, $B$$\rightarrow$$C$ in a chain $A$$\rightarrow$$B$$\rightarrow$$C$), the disconnected project is directly attached to the origin, creating an incorrect edge ($A$$\rightarrow$$C$).
A similar pattern was observed in direction inference: removing explicit matching degraded accuracy more than removing implicit matching, consistent with the precedence of explicit evidence in our design.
Among the clustering criteria, fork relationships, file paths, and file names contributed comparably, while unique functions contributed less but still meaningfully.

\begin{table}[t]
\renewcommand{\tabcolsep}{0.8mm}
    \caption{\label{table:ablation}Ablation study results. Each row removes one component from \sys.}
    \small
    \vspace{-1em}
    \centering
\begin{tabular}{|c|l|r|r|r|r|r|}
\hline
\multicolumn{1}{|c|}{\rule{0in}{2.2ex}\textbf{Phase}}
& \multicolumn{1}{c|}{\rule{0in}{2.2ex}\textbf{Removed component}}
& \multicolumn{1}{c|}{\rule{0in}{2.2ex}\textbf{\#TP}}
& \multicolumn{1}{c|}{\rule{0in}{2.2ex}\textbf{\#FP}}
& \multicolumn{1}{c|}{\rule{0in}{2.2ex}\textbf{\#FN}}
& \multicolumn{1}{c|}{\rule{0in}{2.2ex}\textbf{P(\%)}}
& \multicolumn{1}{c|}{\rule{0in}{2.2ex}\textbf{R(\%)}}
\\\hline\hline
\multicolumn{2}{|c|}{\rule{0in}{2.2ex}\sys (all components)} & \textcolor{tomato}{\textbf{1,535}} & \textcolor{tomato}{\textbf{274}} & \textcolor{tomato}{\textbf{68}} & \textcolor{tomato}{\textbf{84.85}} & \textcolor{tomato}{\textbf{95.76}} \\\hline
\multirow{4}{*}{\rule{0in}{2.2ex}Clustering (P1)} & Fork relationships & 953 & 511 & 650 & 65.10 & 59.45 \\\cline{2-7}
 & File paths & 977 & 514 & 626 & 65.53 & 60.95 \\\cline{2-7}
 & File names & 962 & 511 & 641 & 65.31 & 60.01 \\\cline{2-7}
 & Unique functions & 1,145 & 385 & 458 & 74.84 & 71.43 \\\hline
\multirow{2}{*}{\rule{0in}{2.2ex}Direction (P2)} & Explicit matching & 963 & 503 & 640 & 65.69 & 60.07 \\\cline{2-7}
 & Implicit matching & 1,098 & 391 & 505 & 73.74 & 68.50 \\\hline
\multirow{2}{*}{\rule{0in}{2.2ex}Graph (P3)} & Explicit edges & 849 & 605 & 754 & 58.39 & 52.96 \\\cline{2-7}
 & Implicit edges & 1,077 & 363 & 526 & 74.79 & 67.19 \\\hline
\end{tabular}
\end{table}

\begin{figure}[t]
	
	\begin{center}
		\begin{subfigure}[b]{0.23\textwidth}
			\centering
			\includegraphics[width=\linewidth]{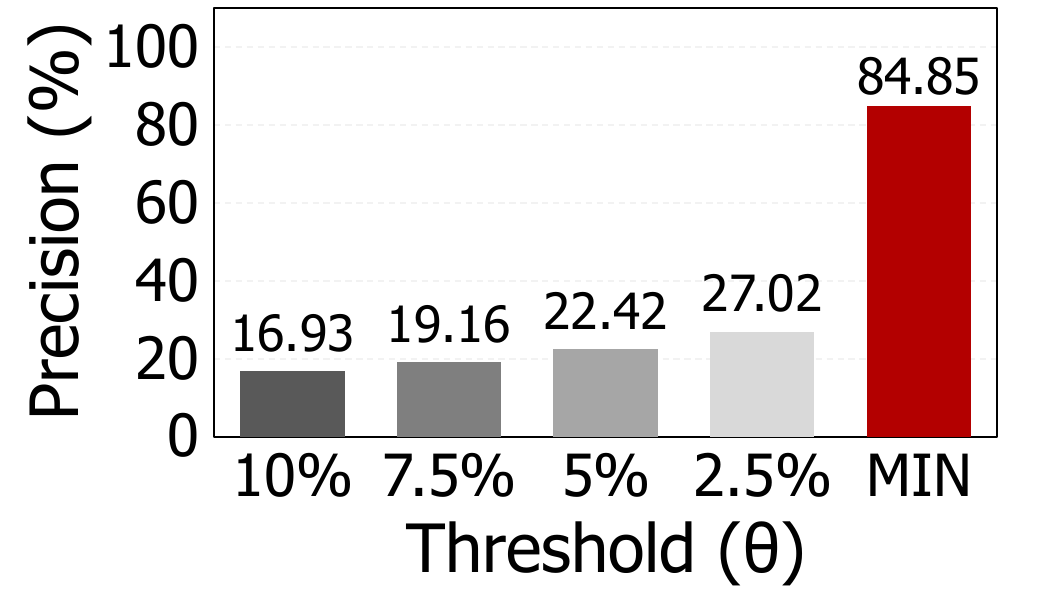}
	
			\vspace{-0.3em}
			\caption{Precision}
			\label{fig:theta_precision}
		\end{subfigure}\hspace{0.0em}%
		\begin{subfigure}[b]{0.23\textwidth}
			\centering
			\includegraphics[width=\linewidth]{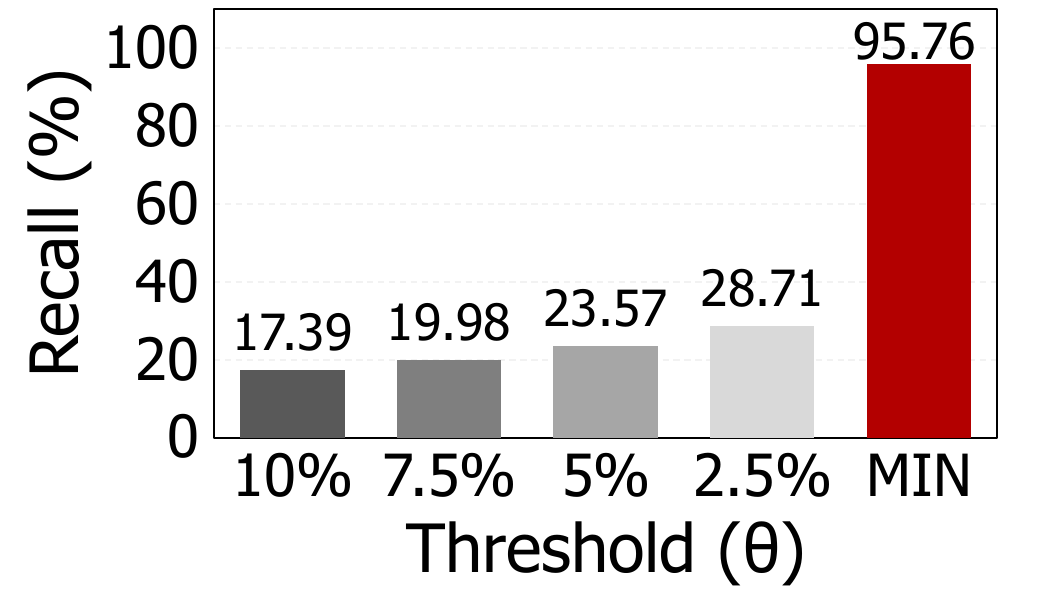}
			
			\vspace{-0.3em}
			\caption{Recall}
			\label{fig:theta_recall}
		\end{subfigure}
		\vspace{-0.8em}
  
		\caption{Effect of $\theta$ on precision and recall.}\label{fig:theta}
	\end{center}
	
	\vspace{-0.83em}
\end{figure}

\begin{figure}[t]
	\begin{center}
		\includegraphics[width=0.85\linewidth]{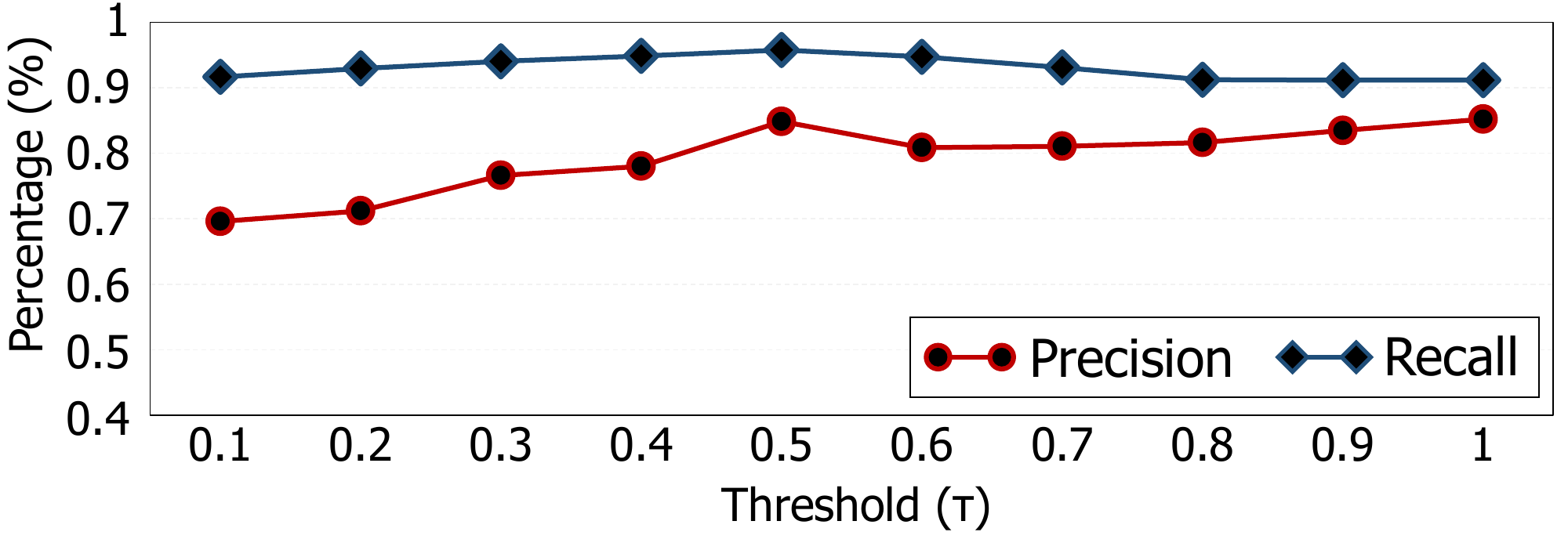}	
		
		\vspace{-0.5em}
		\caption{\label{fig:pr_tau}Effect of $\tau$ on precision and recall.}
	\end{center}
	
	\vspace{-1em}
\end{figure}

\begin{figure}[t]
	\begin{center}
		\includegraphics[width=0.85\linewidth]{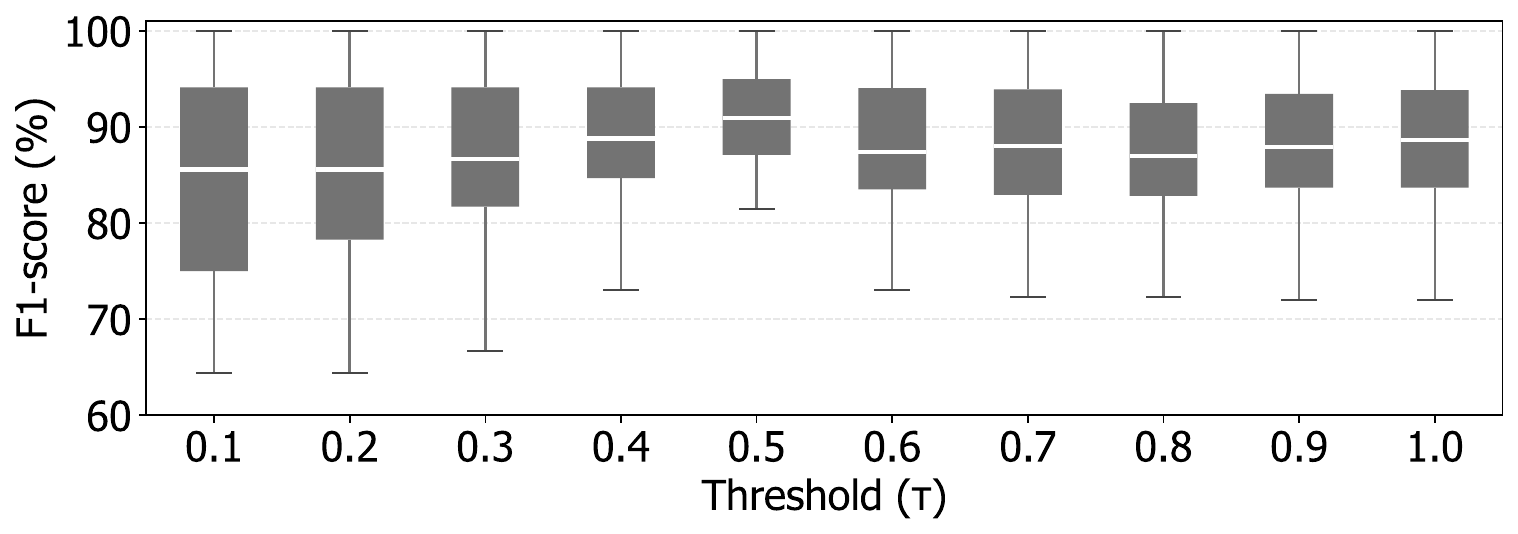}	
		
		\vspace{-0.5em}
		\caption{\label{fig:tau}Effect of $\tau$ on F1-score.}
	\end{center}
	
	\vspace{-1em}
\end{figure}

\begin{figure*}[t]
    \centering
    \begin{subfigure}[b]{0.3\textwidth}
        \centering
        \includegraphics[width=\linewidth]{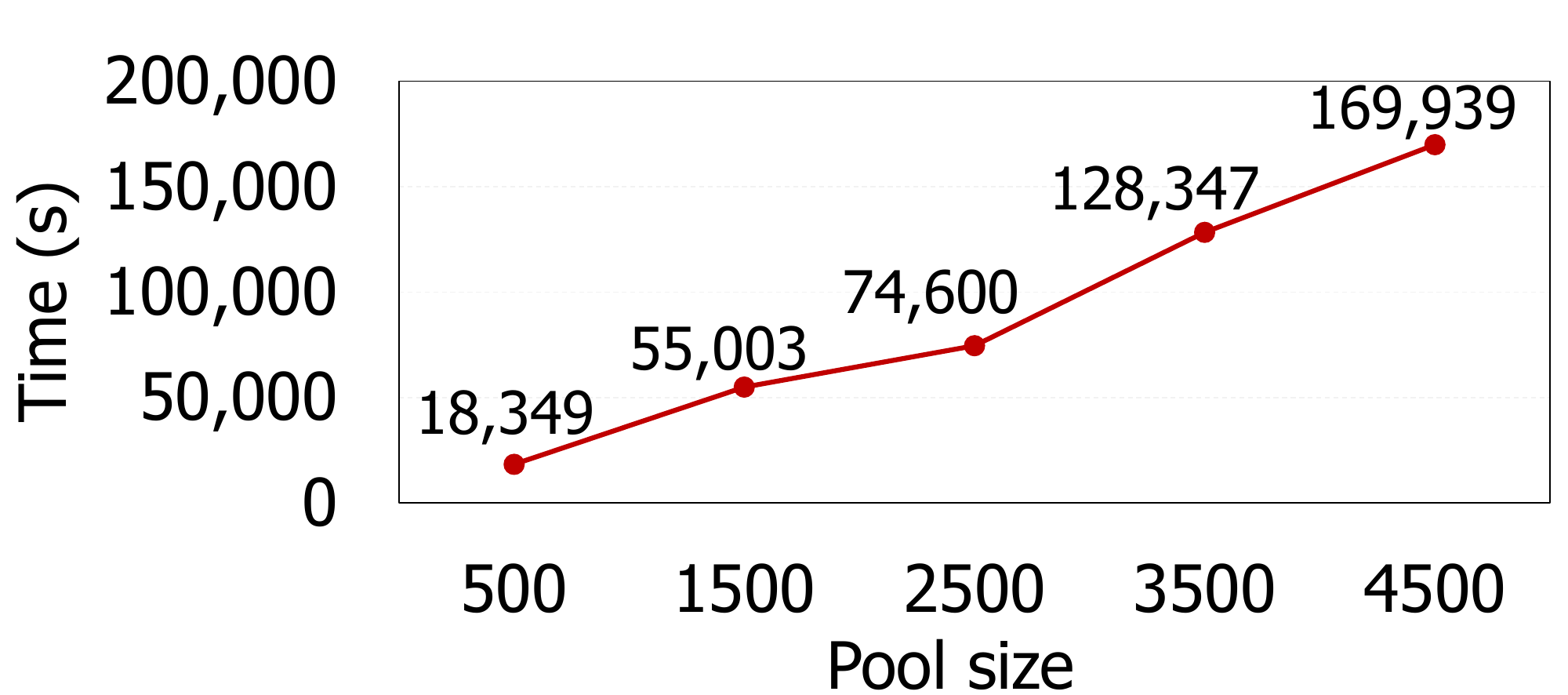}

        \vspace{-0.3em}
        \caption{Pool construction time}
        \label{fig:eff_v0}
    \end{subfigure}\hspace{1em}%
    \begin{subfigure}[b]{0.3\textwidth}
        \centering
        \includegraphics[width=\linewidth]{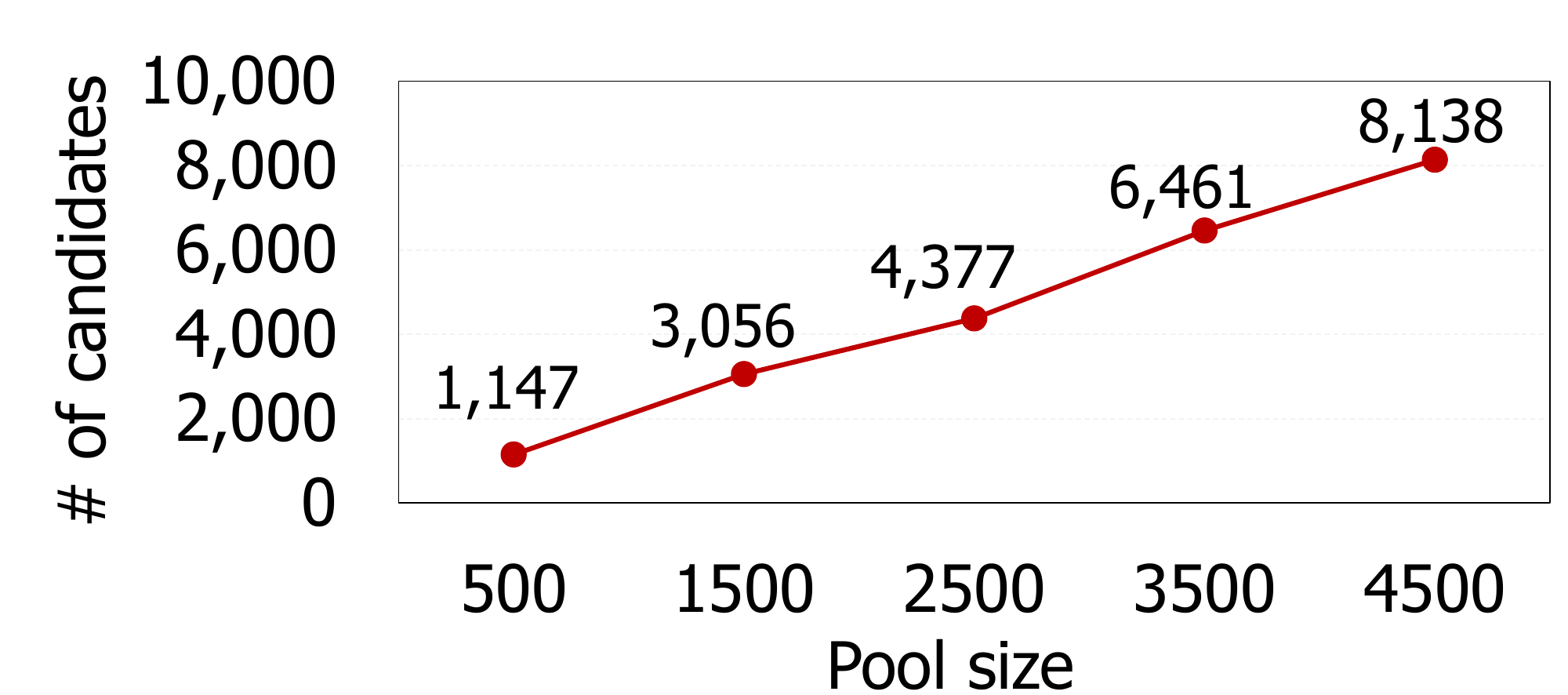}
        
        \vspace{-0.3em}
        \caption{\# of candidates}
        \label{fig:eff_cn}
    \end{subfigure}\hspace{1em}%
    \begin{subfigure}[b]{0.3\textwidth}
        \centering
        \includegraphics[width=\linewidth]{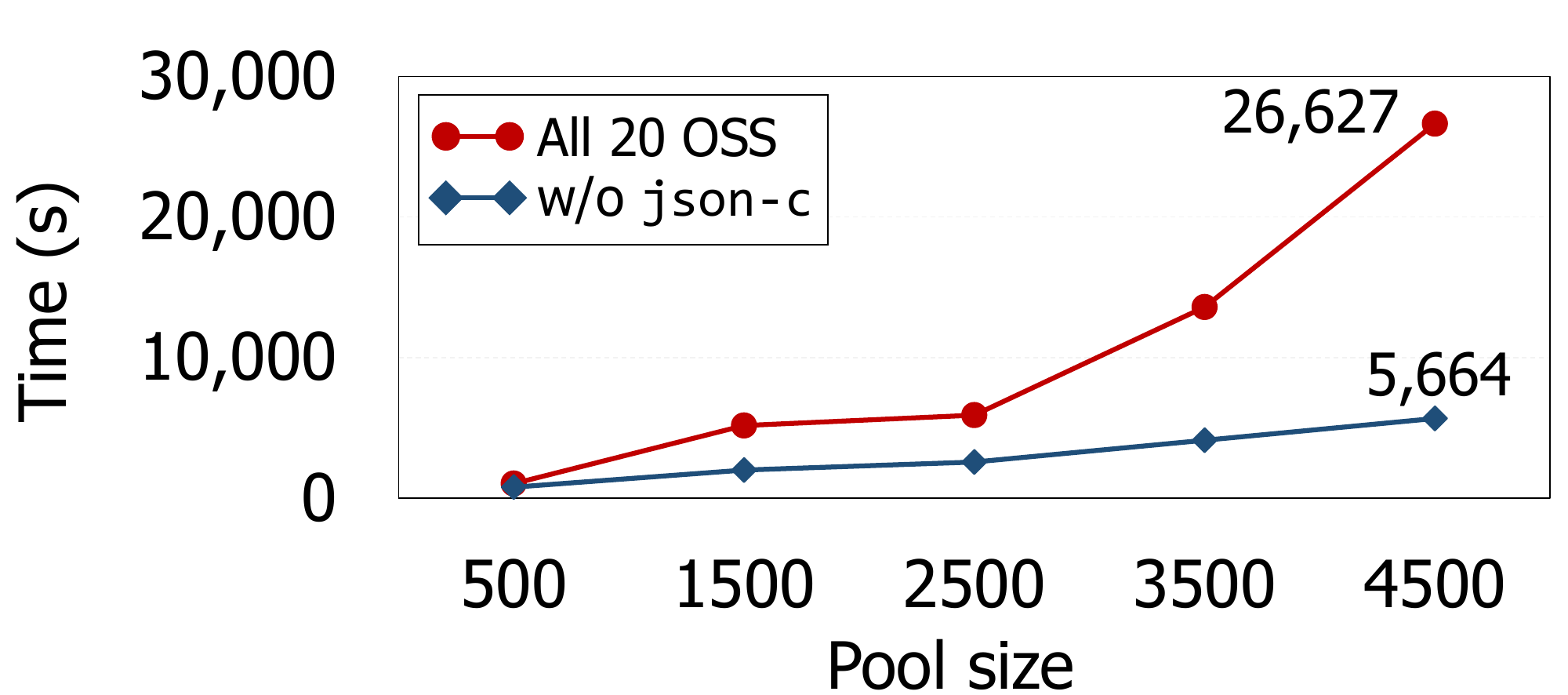}
        
        \vspace{-0.3em}
        \caption{Genealogy identification time}
        \label{fig:eff_ganadi}
    \end{subfigure}
    \vspace{-0.6em}

    \caption{\label{fig:scal}Scalability of \sys with varying pool sizes: (a) pool construction, (b) identified candidates, and (c) identification time for the 20 target projects (\autoref{table:acc}). Even as the pool grows, all measurements increase near-linearly except for \texttt{json-c}.} 
\end{figure*}

\subsubsection{Threshold sensitivity}\label{subsubsec:threshold}
To evaluate the impact of $\theta$ (\autoref{subsec:p1}), we varied it from the minimum setting (\ie, at least one mapped function) to 2.5\%, 5\%, 7.5\%, and 10\%; We limited the maximum value to 10\%, following \cent, as higher values yielded too few candidates.
As shown in \autoref{fig:theta}, both precision and recall dropped significantly as $\theta$ increased. Although a higher $\theta$ was expected to reduce FPs by narrowing the candidate set, it instead increased both FPs (due to missing intermediate projects) and FNs (due to missed TPs). Hence, \sys adopts the minimum $\theta$ as the default.

To assess the impact of $\tau$ (\autoref{subsec:p3}), we fixed $\theta$ at its minimum value and varied $\tau$ from 0.1 to 1.0 in increments of 0.1.
\autoref{fig:pr_tau} shows the precision and recall aggregated over the 20 target OSS projects, and \autoref{fig:tau} shows the distribution of F1-scores across individual origins.
As shown in \autoref{fig:pr_tau}, precision consistently improved as $\tau$ increased, as higher thresholds filter out weak similarity edges.
Recall, however, peaked at $\tau = 0.5$ and degraded in both directions: higher $\tau$ values discarded genuine reuse relationships, while lower values admitted noisy edges that displaced correct ones during graph construction, increasing FNs as well as FPs.
\autoref{fig:tau} shows a trend at the F1 level: \mbox{$\tau = 0.5$} achieved the highest average F1-score with relatively low variance, and the F1-scores remained stable across a wide range of $\tau$ (0.3 to 1.0), indicating that the accuracy of \sys is robust to the choice of $\tau$.
Thus, we selected $\tau = 0.5$ as the default.
Note that $\theta$ and $\tau$ are the only tunable parameters of \sys; the other design elements (\eg, the clustering criteria and graph construction rules) are rule-based and require no tuning.

\subsection{Performance and Scalability}\label{subsec:performance}

\subsubsection{Performance} To evaluate performance, we measured the time required for \sys to construct the software pool and to build the OSS reuse genealogies.
First, constructing the software pool (\autoref{subsubsec:pool}) took 20.8 hours in our environment. However, this is a one-time preprocessing step and does not impact run-time performance.
Next, identifying the reuse genealogy for the 20 target OSS projects (excluding database I/O time) took 314.29 s per project on average, with a median of 86.83 s.
The gap between the average and median is caused by \texttt{json-c}, which yielded 644 candidates and required approximately one hour to analyze, because the runtime is largely determined by the number of candidates.
For all other OSS projects (fewer than 200 candidates), the average execution time was less than one minute.
Even under the most aggressive setting (\ie, the minimum $\theta$), \sys took only 148.79 s on average per project (excluding outliers), demonstrating its efficiency.

\subsubsection{Scalability}\label{subsubsec:scal}
To evaluate how \sys scales with the software pool, we varied the pool size from 500 to 4,500 repositories and measured, for the 20 target OSS projects, (1) pool construction time, (2) the number of identified candidates, and (3) genealogy identification time. \autoref{fig:scal} presents the measurement results.

Pool construction time grew linearly with the pool size (9.3 times for a nine-fold larger pool), as every step before candidate identification handles each repository independently.

The total number of candidates grew more slowly than the pool (7.1 times under the nine-fold growth); because only the projects that share functions with the origin become candidates, expanding the pool does not proportionally increase the comparison targets.

Identification time did not exhibit quadratic growth: excluding \texttt{json-c}, it grew 8.1 times, closely tracking the number of candidates rather than its quadratic bound, because clustering partitions candidates into smaller groups and confines pairwise comparisons within each cluster.
The only exception was \texttt{json-c}, whose identification time grew 83.9 times; \texttt{json-c} is reused so widely that it yielded far more candidates than any other origin (1,406 at the largest pool, compared to at most 523 for the others), incurring the quadratic pairwise cost.
Even in this case, however, the analysis remains tractable, taking approximately one hour in the 2,500-repository pool as a one-time analysis per origin.

\subsubsection{Complexity analysis}
To examine the scalability of \sys, we analyze the algorithmic complexity of each phase.

Let $n$ denote the pool size and $c$ denote the number of candidates for an origin identified by \sys.
First, pool construction costs $O(n)$, as it processes the functions of each repository exactly once, independently of the others.
Candidate identification also costs $O(n)$: it screens the pool by matching the origin's function hashes against each repository's function set, where each check is a constant-time hash lookup.
The remaining phases are independent of $n$ and depend only on $c$: clustering costs $O(c^2)$, as it examines the four clustering criteria for every pair of candidates.
Reuse inference costs $O(\sum_i |C_i|^2)$, where $C_i$ denotes the resulting clusters, because pairwise similarity and direction comparisons are confined within each cluster; note that $\sum_i |C_i|^2 \le c^2$, with equality only when all candidates fall into a single cluster.
Graph construction is linear in the number of inferred edges, as rules R1 to R4 require only a constant number of comparisons per edge.

Our scalability experiments (\autoref{subsubsec:scal}) are consistent with this analysis: as the pool grew, the number of candidates grew sublinearly, and the clustering kept the pairwise cost far below its quadratic bound, so that the overall identification time grew roughly in line with the pool size, except for \texttt{json-c} described above. 
For pools beyond our experimental scale, the same behavior is expected: except for such rare origins that are almost universally reused, the overall cost grows near-linearly with the pool, indicating that \sys scales to much larger pools.

\begin{table}[t]
\centering
\renewcommand{\tabcolsep}{0.4mm}	
    \caption{\label{table:vul_res}Vulnerability detection results with and without genealogy (432 TPs identified via manual analysis).}
    \small
    \vspace{-0.6em}
\begin{tabular}{|c|c|c|c|c|l|l|l|}
\hline
\rule{0in}{2.2ex}\textbf{Tool} 
& \textbf{Setting}
& \textbf{\#TP} 
& \textbf{\#FP} 
& \textbf{\#FN} 
& \multicolumn{1}{|c|}{\textbf{P(\%)}} 
& \multicolumn{1}{|c|}{\textbf{R(\%)}} 
& \multicolumn{1}{|c|}{\textbf{F1(\%)}} \\ \hline\hline
\multirow{2}{*}{\begin{tabular}[c]{@{}c@{}}\rule{0in}{2.2ex}\textbf{FIRE}\end{tabular}}&\rule{0in}{2.2ex}Baseline& 420  & 177  & 12   & 70.35 & 97.22 & 81.63  \\
&Genealogy  & 382  & 74   & 50 & 83.77 {\footnotesize \textcolor{asegreen}{(+13.42)}}& 88.43 {\footnotesize \textcolor{purple}{(-8.79)}}& 86.04 {\footnotesize \textcolor{asegreen}{(+4.41)}} \\\hline
\multirow{2}{*}{\begin{tabular}[c]{@{}c@{}}\rule{0in}{2.2ex}\textbf{VUDDY}\end{tabular}}&\rule{0in}{2.2ex}Baseline & 146  & 93   & 286  & 61.09 & 33.80 & 43.52  \\
&Genealogy  & 115  & 38   & 317  & 75.16 {\footnotesize \textcolor{asegreen}{(+14.07)}}& 26.62 {\footnotesize \textcolor{purple}{(-7.18)}}& 39.32  {\footnotesize \textcolor{purple}{(-4.20)}}\\\hline
\end{tabular}
\end{table}

\subsection{Application: Vulnerability Detection}\label{subsec:app}

Next, we evaluate how \sys can strengthen supply chain security via vulnerability detection.

\subsubsection{Methodology}
We evaluate genealogy-driven vulnerability detection by applying two existing tools,
VUDDY~\cite{kim2017vuddy} and FIRE~\cite{feng2024fire},
under two settings: 
with and without genealogy information.

In \autoref{subsec:accuracy}, \sys identified the genealogies of 20 OSS projects, which collectively include 957 unique software projects (\ie, downstreams).
Using the NVD JSON feed, we collected CVEs that provide patches as \texttt{GitHub} commits~\cite{woo2022movery, woo2023v1scan} across these projects, obtaining 1,850 unique CVEs.

Based on this, we conduct the following two experiments:

\vspace{0.2em}
\begin{itemize}
\setlength\itemsep{0.1em}
\item \textbf{Baseline (no genealogy).} We apply VUDDY and FIRE without leveraging genealogy information by mapping each of the 957 software projects to all 1,850 CVEs previously reported in the collected projects.

\item \textbf{Genealogy-driven approach.}  
When analyzing a given software project, we restrict the vulnerability search space using genealogy information by tracing only CVEs previously reported in projects within the same genealogy, with approximately 200 unique CVEs per genealogy on average.
\end{itemize}
\vspace{0.3em}

We exclude vulnerabilities reported by software that were included in \sys's FPs (see \autoref{subsec:accuracy}).
This exclusion isolates the impact of genealogy accuracy on vulnerability management, as including vulnerabilities from unrelated projects would introduce noise unrelated to genealogy quality.
The detection results of both tools were manually analyzed and categorized into TPs and FPs by the two researchers who conducted the accuracy evaluation.

To further leverage genealogy information, we examined the potential FNs of the two tools as follows. First, we selected a target vulnerability from one project within a genealogy. Second, we examined possible propagation in a structured order (downstream, upstream, then other related projects), analyzing the security patch to identify vulnerable functions, defined as those containing code lines removed by the patch~\cite{kim2017vuddy, woo2022movery, woo2023v1scan, feng2024fire}. Third, for each project in the genealogy, we verified whether the vulnerable function had been reused and whether the fix had been applied; when the reused code remained unpatched, we attempted to trigger the vulnerability using the original Proof-of-Concept when available.

\subsubsection{Result analysis}
\autoref{table:vul_res} presents the results.
The baseline identified more TPs, however, this gap is not a fundamental limitation; it is a direct consequence of \sys's genealogy FNs. Because the genealogy-driven approach restricts its search space to projects within a given genealogy, vulnerabilities originating from missed downstreams fall outside its search space, and are detectable only by the exhaustive baseline.

Despite this, the genealogy-driven approach demonstrates three notable strengths.
First, it \textbf{improves precision} by restricting the analysis scope to evolutionarily and structurally related projects. In this setting, FIRE's precision increases from 70.35\% to 83.77\%, and VUDDY's from 61.09\% to 75.16\%. This reduction in FPs lowers validation overhead and improves practical usability.

Second, the genealogy-driven approach achieves comparable detection performance while \textbf{relying on a much smaller CVE dataset}. 
It uses only 11\% on average (200 per genealogy \textit{vs.} 1,850), yet the F1-score increases by 4.41\% for FIRE and decreases by only 4.2\% for VUDDY, 
indicating that similar detection performance can be maintained without an exhaustive vulnerability dataset while reducing analysis overhead.

Last, the genealogy information provides explicit propagation paths across related software projects, \textbf{offering structural context on how vulnerabilities are inherited}. This enables more systematic manual validation and investigation. In practice, using the aforementioned manual procedure, two experts analyzed vulnerabilities across 20 genealogies in under one hour and identified 12 vulnerabilities that were not detected by FIRE and VUDDY.

\begin{figure}[t]
	\begin{center}
		\includegraphics[width=0.85\linewidth]{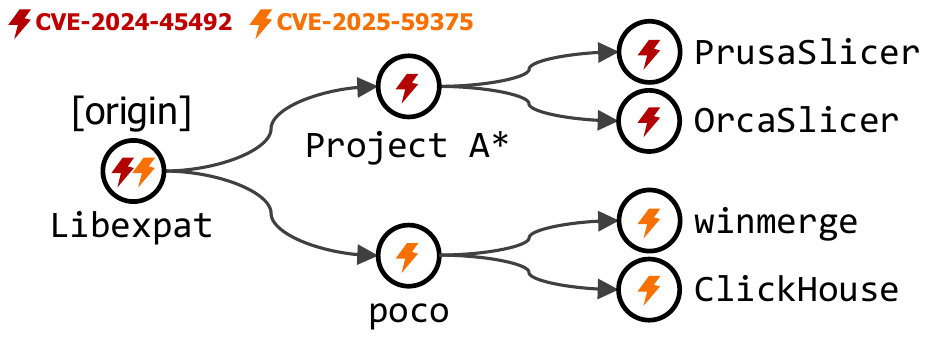}	
		
		\caption{\label{fig:app_graph}
        OSS reuse genealogy graph for \texttt{Libexpat}, showing two vulnerabilities propagating along distinct paths. We reported them to all repositories in the graph. Project A remains unpatched and is anonymized to prevent misuse.
        }
	\end{center}
	
	\vspace{-0.5em}
\end{figure}

\subsubsection{Responsible disclosure}
During the experiment, we identified 48 triggerable, unpatched propagated vulnerabilities, including 12 vulnerabilities manually detected.
We reported all vulnerabilities to the respective teams or maintainers.
As of March 2026, 23 vulnerabilities have been patched, including those in widely used software such as \texttt{F-Stack}, \texttt{Tendis}, and \texttt{Wireshark}. 
One CVE has been assigned for our reported findings (CVE-2025-26269 in \texttt{Dragonfly}). 

\vspace{-0.05em}
\PP{Case study}
\texttt{Libexpat} reported CVE-2024-45492 (an integer overflow vulnerability; severity: Critical) in 2024 and CVE-2025-59375 (an unbounded resource allocation vulnerability; severity: High) in 2025.
Upon examining the genealogy of \texttt{Libexpat}, we identified previously undiscovered vulnerability propagation paths, as illustrated in \autoref{fig:app_graph}. We reported these findings, which were subsequently patched.
In particular, downstream projects such as \texttt{WinMerge} and \texttt{ClickHouse}, which incorporate \texttt{Libexpat} via \texttt{poco}, made it difficult to analyze the propagation of these vulnerabilities through conventional means. However, genealogy-based vulnerability management enabled systematic tracing of the propagation paths, demonstrating its effectiveness in uncovering previously unidentified vulnerabilities.
\section{Discussion}\label{sec:diss}
\vspace{-2px}
\vspace{-1px}
\PP{Threats to validity}
In the experiment, we collected 2,500 repositories to build a pool and analyzed the reuse genealogy of 20 target OSS. The results identified over 1,600 reuse relationships; however, this remains insufficient to reflect the overall software environment.
In addition, due to the absence of ground truth, we manually analyzed \sys's results. The two validators independently agreed on 98\% of the results. For the remaining 2\%, they reviewed the relevant evidence and resolved their disagreements through discussion. Because the senior validator's judgment was adopted in most of these cases, the resulting labels may still contain human error or subjective bias.
Moreover, due to the absence of existing research with nearly the same purpose as ours, we indirectly compared \sys with \cn and \vzfinder.
We do not intend to diminish the existing studies, but rather to demonstrate that \sys's algorithm is effective for reuse genealogy analysis.
\vspace{1px}

\PP{Limitations and future work}
First, \sys operates in environments where C/C++ source code is available. 
Next, \sys may fail to infer reuse directionality when explicit evidence is unavailable. When neither fork relationships nor file-path signals exist, \sys falls back on implicit matching, which relies on code birth dates. If timestamps are inconsistent, date-based comparison may yield incorrect directions, and when path-based comparison also provides no signal, directionality inference may fail entirely. Although explicit matching takes strict precedence and mitigates most such cases, we acknowledge this residual limitation. We plan to mitigate it by considering additional features (\eg, incorporating natural language processing into \texttt{README} files).
Moreover, there are reuse relationships that cannot be identified by the pivotal function-based approach. We plan to consider incorporating structural information and various metadata files. 
Finally, \sys focuses only on analyzing reuse genealogy, while vulnerability identification requires the use of other tools or human analysis. We plan to devise a framework that can perform comprehensive vulnerability management.

\vspace{1px}
\PP{Language extensibility}
Although \sys targets C/C++, its methodology is largely language-agnostic, as it relies on function-level code and metadata that are available in most programming languages. Nevertheless, applying \sys to other ecosystems would require accounting for their ecosystem-specific characteristics. For example, in ecosystems with well-established package managers (\eg, Java), code reuse primarily occurs through declared dependencies, leaving fewer copy-based traces to analyze. At the same time, these ecosystems often provide richer explicit metadata, such as dependency manifests, which could be incorporated into \sys to improve genealogy accuracy.
\vspace{0.1em}
\section{Related Work}\label{sec:related}

\PP{Software composition analysis (SCA).} 
Several approaches attempt to identify reused third-party libraries in software codebases (\eg, \cite{  na2024cneps, jiang2024binaryai,kim2025debun,wu2024vision, yang2026sbridge}).
For example, some approaches (\eg, \cite{woo2021centris,wu2023ossfp,jiang2023third}) identify OSS components by adopting Locality Sensitive Hashing algorithms at the function level.
\textsc{Cneps}~\cite{na2024cneps} analyzes dependencies between components based on function call-based modules.
\textsc{BinaryAI}~\cite{jiang2024binaryai} identifies OSS in binary files by combining a transformer model and link-time locality.
However, these approaches can only identify the origins, without capturing the ancestor–descendant relationships in the OSS genealogy. 

\PP{Origin and code evolution analysis.} 
Some studies have analyzed the origin and evolution of source code
(\eg, \cite{inoue2012does,godfrey2008past,steidl2014incremental, kanda2013extraction, hata2022software, kim2005empirical,jahanshahi2025beyond}).
Inoue \etal~\cite{inoue2012does} analyzed the genealogy of code clones focused on tracking clones within a single project, while Steidl \etal~\cite{steidl2014incremental} provided methods for tracking code history during project version updates. However, neither study addressed relationships across different projects.
Kanda \etal~\cite{kanda2013extraction} used the longest common subsequence to compare the number of similar files and built a product evolution tree, but their analysis failed to capture propagation direction or change history.  
Hata \etal~\cite{hata2022software} aimed to uncover clone-and-own reuse relationships across 4,592 projects through n-gram similarity, but failed to closely identify how component reuse propagates.
Although most prior work has focused on the evolution within a single project, our approach traces how components propagate across multiple projects throughout the ecosystem.

\PP{Code clone detection.}  
Many studies aim to identify code clones 
(\eg, \cite{gode2009incremental, kamiya2002ccfinder, sajnani2016sourcerercc, shan2023gitor, yu2025multiple}),
which can be leveraged to identify reuse genealogies. 
For example, \textsc{Gitor}~\cite{shan2023gitor} uses a global code graph, while Yu \etal~\cite{yu2025multiple} combine pruned ASTs with a Siamese transformer.
Although \sys builds on code similarity, it differs fundamentally from clone detection: existing clone detection approaches answer a symmetric question (whether two fragments are similar), whereas \sys answers an asymmetric one (which project reused code from which, through what path). To this end, \sys introduces components absent in clone detection: pivotal function-based clustering scoped to origin-derived regions, direction inference via explicit evidence and majority voting, and priority-based graph construction (R1-R4).
\section{Conclusion}\label{sec:conclusion}
As the reuse of OSS becomes widespread and software supply chains become more complex, identifying reuse genealogies is essential from the perspective of supply chain security. We propose \sys, an approach that accurately identifies OSS reuse genealogies by leveraging the concepts of pivotal functions and multi-criteria clustering. \sys achieved high accuracy even in complex reuse ecosystems and was able to uncover previously unpatched vulnerabilities, demonstrating its practical value in real-world supply chain security. By leveraging the results of \sys, developers can perform comprehensive component management and ultimately contribute to building more secure software supply chains.

\section*{Data Availability}
The artifacts of \sys are publicly available on Zenodo~\cite{ganadi-artifact} 
(\url{https://doi.org/10.5281/zenodo.21768585}) and GitHub (\url{https://github.com/KIMDONGYEON00/GANADI}).

\section*{Acknowledgments}
This work was supported by the Institute of Information \& Communications Technology Planning
\& Evaluation (IITP) grant funded by the Korea government (MSIT) (No.RS-2024-00440780, Development of Automated SBOM and VEX Verification Technologies for Securing Software Supply
Chains), ICT Creative Consilience Program (IITP-2026-RS-2020-II201819, 10\%), and the National Research Foundation of Korea (NRF) grant funded by the Korea government
(MSIT) (RS-2025-00517788, Research on Intelligent SBOM Generation and Automated Vulnerability
Analysis through Multi-level Code Analysis).

\bibliographystyle{ACM-Reference-Format}
\bibliography{reference}

@Misc{ganadi-artifact,
  author       = {Kim, Dongyeon and Woo, Seunghoon and Lee, Heejo},
  title        = {Artifact for `GANADI: Uncovering C/C++ OSS Reuse Genealogies via Pivotal Function-Based Clustering to Enhance Supply Chain Security'},
  howpublished = {Zenodo},
  year         = {2026},
  doi          = {10.5281/zenodo.21768585},
}

@article{jahanshahi2025beyond,
  title={{Beyond Dependencies: The Role of Copy-Based Reuse in Open Source Software Development}},
  author={Jahanshahi, Mahmoud and Reid, David and Mockus, Audris},
  journal={ACM Transactions on Software Engineering and Methodology},
  volume={34},
  number={8},
  pages={1--49},
  year={2025},
  publisher={ACM New York, NY},
  doi = {{10.1145/3715907}}
}

@article{yang2026sbridge,
  title={{SBridge: Identifying Source-to-Binary Function Similarity via Cross-Domain Control Block Matching}},
  author={Yang, Heedong and Lee, Jeongwoo and Yun, Hajin and Woo, Seunghoon},
  journal={Proceedings of the ACM on Software Engineering},
  volume={3},
  number={FSE},
  pages={1381--1403},
  year={2026},
  publisher={ACM New York, NY, USA},
  doi = {10.1145/3797090}
}

@inproceedings{kim2025debun,
  title={{Debun: Detecting Bundled JavaScript Libraries on Web using Property-Order Graphs}},
  author={Kim, Seojin and Park, Sungmin and Park, Jihyeok},
  booktitle={2025 40th IEEE/ACM International Conference on Automated Software Engineering (ASE)},
  pages={78--90},
  year={2025},
  organization={IEEE},
  doi = {10.1109/ASE63991.2025.00015}
}

@inproceedings{wu2024vision,
  title={{Vision: Identifying Affected Library Versions for Open Source Software Vulnerabilities}},
  author={Wu, Susheng and Wang, Ruisi and Huang, Kaifeng and Cao, Yiheng and Song, Wenyan and Zhou, Zhuotong and Huang, Yiheng and Chen, Bihuan and Peng, Xin},
  booktitle={Proceedings of the 39th IEEE/ACM International Conference on Automated Software Engineering},
  pages={1447--1459},
  year={2024},
  doi = {10.1145/3691620.3695516}
}

@article{camp2021sbom,
  title={{SBOM Vulnerability Assessment \& Corresponding Requirements}},
  author={Camp, L Jean and Andalibi, Vafa},
  journal={NTIA Response to Notice and Request for Comments on Software Bill of Materials Elements and Considerations},
  year={2021}
}

@inproceedings{o2023impacts,
  title={{Impacts of Software Bill of Materials (SBOM) Generation on Vulnerability Detection}},
  author={O'Donoghue, Eric and Boles, Brittany and Izurieta, Clemente and Reinhold, Ann Marie},
  booktitle={Proceedings of the 2024 Workshop on Software Supply Chain Offensive Research and Ecosystem Defenses},
  pages={67--76},
  year={2023},
  doi = {10.1145/3689944.3696164}
}

@inproceedings{xiao2020mvp,
  title={{MVP: Detecting Vulnerabilities using Patch-Enhanced Vulnerability Signatures}},
  author={Xiao, Yang and Chen, Bihuan and Yu, Chendong and Xu, Zhengzi and Yuan, Zimu and Li, Feng and Liu, Binghong and Liu, Yang and Huo, Wei and Zou, Wei and others},
  booktitle={29th USENIX Security Symposium (USENIX Security 20)},
  pages={1165--1182},
  year={2020}
}

@online{cisa2025sbom,
  title        = {{2025 Minimum Elements for a Software Bill of Materials (SBOM)}},
  organization = {CISA},
  year         = {2025},
  url          = {https://www.ntia.gov/page/software-bill-materials},
  key     = {CISA},
  urldate      = {2026-03-04}
}

@inproceedings{woo2023v1scan,
  title={{V1SCAN: Discovering 1-day Vulnerabilities in Reused C/C++ Open-source Software Components Using Code Classification Techniques}},
  author={Woo, Seunghoon and Choi, Eunjin and Lee, Heejo and Oh, Hakjoo},
  booktitle={32nd USENIX Security Symposium (USENIX Security 23)},
  pages={6541--6556},
  year={2023}
}

@inproceedings{jiang2024binaryai,
  title={{BinaryAI: Binary Software Composition Analysis via Intelligent Binary Source Code Matching}},
  author={Jiang, Ling and An, Junwen and Huang, Huihui and Tang, Qiyi and Nie, Sen and Wu, Shi and Zhang, Yuqun},
  booktitle={Proceedings of the IEEE/ACM 46th International Conference on Software Engineering},
  pages={1--13},
  year={2024},
  doi = {10.1145/3597503.3639100}
}

@article{pedregosa2011scikit,
  title={{Scikit-learn: Machine Learning in Python}},
  author={Pedregosa, Fabian and Varoquaux, Ga{\"e}l and Gramfort, Alexandre and Michel, Vincent and Thirion, Bertrand and Grisel, Olivier and Blondel, Mathieu and Prettenhofer, Peter and Weiss, Ron and Dubourg, Vincent and others},
  journal={the Journal of machine Learning research},
  volume={12},
  pages={2825--2830},
  year={2011},
  publisher={JMLR. org}
}

@inproceedings{yu2025multiple,
  title={{A Multiple Representation Transformer with Optimized Abstract Syntax Tree for Efficient Code Clone Detection}},
  author={Yu, Tianchen and Yuan, Li and Lin, Liannan and He, Hongkui},
  booktitle={2025 IEEE/ACM 47th International Conference on Software Engineering (ICSE)},
  pages={587--587},
  year={2025},
  organization={IEEE Computer Society},
  doi = {10.1109/ICSE55347.2025.00050}
}

@inproceedings{steidl2014incremental,
  title={{Incremental Origin Analysis of Source Code Files}},
  author={Steidl, Daniela and Hummel, Benjamin and Juergens, Elmar},
  booktitle={Proceedings of the 11th Working Conference on Mining Software Repositories},
  pages={42--51},
  year={2014}
}

@inproceedings{vendome2015and,
  title={{When and why developers adopt and change software licenses}},
  author={Vendome, Christopher and Linares-V{\'a}squez, Mario and Bavota, Gabriele and Di Penta, Massimiliano and German, Daniel M and Poshyvanyk, Denys},
  booktitle={2015 IEEE international conference on software maintenance and evolution (ICSME)},
  pages={31--40},
  year={2015},
  organization={IEEE}
}

@article{williams2025research,
  title={{Research Directions in Software Supply Chain Security}},
  author={Williams, Laurie and Benedetti, Giacomo and Hamer, Sivana and Paramitha, Ranindya and Rahman, Imranur and Tamanna, Mahzabin and Tystahl, Greg and Zahan, Nusrat and Morrison, Patrick and Acar, Yasemin and others},
  journal={ACM Transactions on Software Engineering and Methodology},
  volume={34},
  number={5},
  pages={1--38},
  year={2025},
  publisher={ACM New York, NY}
}

@inproceedings{di2010exploratory,
  title={{An Exploratory Study of the Evolution of Software Licensing}},
  author={Di Penta, Massimiliano and German, Daniel M and Gu{\'e}h{\'e}neuc, Yann-Ga{\"e}l and Antoniol, Giuliano},
  booktitle={Proceedings of the 32nd ACM/IEEE International Conference on Software Engineering-Volume 1},
  pages={145--154},
  year={2010}
}

@inproceedings{liu2022demystifying,
  title={{Demystifying the Vulnerability Propagation and Its Evolution via Dependency Trees in the NPM Ecosystem}},
  author={Liu, Chengwei and Chen, Sen and Fan, Lingling and Chen, Bihuan and Liu, Yang and Peng, Xin},
  booktitle={Proceedings of the 44th International Conference on Software Engineering},
  pages={672--684},
  year={2022}
}

@article{hata2022software,
  title={{Software Supply Chain Map: How Reuse Networks Expand}},
  author={Hata, Hideaki and Ishio, Takashi},
  journal={arXiv preprint arXiv:2204.06531},
  year={2022}
}

@inproceedings{wu2024large,
  title={{A Large-Scale Empirical Study of Open Source License Usage: Practices and Challenges}},
  author={Wu, Jiaqi and Bao, Lingfeng and Yang, Xiaohu and Xia, Xin and Hu, Xing},
  booktitle={Proceedings of the 21st International Conference on Mining Software Repositories},
  pages={595--606},
  year={2024}
}

@inproceedings{huang2024vmud,
  title={{VMUD: Detecting Recurring Vulnerabilities with Multiple Fixing Functions via Function Selection and Semantic Equivalent Statement Matching}},
  author={Huang, Kaifeng and Lu, Chenhao and Cao, Yiheng and Chen, Bihuan and Peng, Xin},
  booktitle={Proceedings of the 2024 on ACM SIGSAC Conference on Computer and Communications Security},
  pages={3958--3972},
  year={2024},
  doi = {10.1145/3658644.3690372}
}

@inproceedings{kwon2021octopocs,
  title={{OCTOPOCS: Automatic Verification of Propagated Vulnerable Code Using Reformed Proofs of Concept}},
  author={Kwon, Seongkyeong and Woo, Seunghoon and Seong, Gangmo and Lee, Heejo},
  booktitle={2021 51st Annual IEEE/IFIP International Conference on Dependable Systems and Networks (DSN)},
  pages={174--185},
  year={2021},
  organization={IEEE}
}

@inproceedings{feng2024fire,
  title={{FIRE: Combining Multi-Stage Filtering with Taint Analysis for Scalable Recurring Vulnerability Detection}},
  author={Feng, Siyue and Wu, Yueming and Xue, Wenjie and Pan, Sikui and Zou, Deqing and Liu, Yang and Jin, Hai},
  booktitle={33rd USENIX Security Symposium (USENIX Security 24)},
  pages={1867--1884},
  year={2024}
}

@inproceedings{woo2021v0finder,
  title={{V0Finder: Discovering the Correct Origin of Publicly Reported Software Vulnerabilities}},
  author={Woo, Seunghoon and Lee, Dongwook and Park, Sunghan and Lee, Heejo and Dietrich, Sven},
  booktitle={30th USENIX Security Symposium (USENIX Security 21)},
  pages={3041--3058},
  year={2021}
}

@inproceedings{woo2022movery,
	title={{MOVERY: A Precise Approach for Modified Vulnerable Code Clone Discovery from Modified Open-Source Software Components}},
	author={Woo, Seunghoon and Hong, Hyunji and Choi, Eunjin and Lee, Heejo},
	booktitle={Proceedings of the 31st USENIX Security Symposium (Security)},
	pages={3037--3053},
	year={2022}
}

@inproceedings{na2024cneps,
  title={{CNEPS: A Precise Approach for Examining Dependencies among Third-Party C/C++ Open-Source Components}},
  author={Na, Yoonjong and Woo, Seunghoon and Lee, Joomyeong and Lee, Heejo},
  booktitle={Proceedings of the IEEE/ACM 46th International Conference on Software Engineering},
  pages={1--12},
  year={2024},
  doi = {10.1145/3597503.3639209}
}

@inproceedings{inoue2012does,
  title={{Where Does This Code Come From and Where Does It Go?—Integrated Code History Tracker for Open Source Systems}},
  author={Inoue, Katsuro and Sasaki, Yusuke and Xia, Pei and Manabe, Yuki},
  booktitle={2012 34th International Conference on Software Engineering (ICSE)},
  pages={331--341},
  year={2012},
  organization={IEEE}
}

@inproceedings{kim2017vuddy,
  title={{VUDDY: A Scalable Approach for Vulnerable Code Clone Discovery}},
  author={Kim, Seulbae and Woo, Seunghoon and Lee, Heejo and Oh, Hakjoo},
  booktitle={2017 IEEE symposium on security and privacy (SP)},
  pages={595--614},
  year={2017},
  organization={IEEE},
  doi = {10.1109/SP.2017.62}
}

@inproceedings{nguyen2013study,
  title={{A Study of Repetitiveness of Code Changes in Software Evolution}},
  author={Nguyen, Hoan Anh and Nguyen, Anh Tuan and Nguyen, Tung Thanh and Nguyen, Tien N and Rajan, Hridesh},
  booktitle={2013 28th IEEE/ACM International Conference on Automated Software Engineering (ASE)},
  pages={180--190},
  year={2013},
  organization={IEEE}
}

@inproceedings{kim2005empirical,
  title={{An Empirical Study of Code Clone Genealogies}},
  author={Kim, Miryung and Sazawal, Vibha and Notkin, David and Murphy, Gail},
  booktitle={Proceedings of the 10th European software engineering conference held jointly with 13th ACM SIGSOFT international symposium on Foundations of software engineering},
  pages={187--196},
  year={2005}
}

@inproceedings{kanda2013extraction,
  title={{Extraction of Product Evolution Tree from Source Code of Product Variants}},
  author={Kanda, Tetsuya and Ishio, Takashi and Inoue, Katsuro},
  booktitle={Proceedings of the 17th International Software Product Line Conference},
  pages={141--150},
  year={2013}
}

@inproceedings{godfrey2008past,
  title={{The Past, Present, and Future of Software Evolution}},
  author={Godfrey, Michael W and German, Daniel M},
  booktitle={2008 Frontiers of Software Maintenance},
  pages={129--138},
  year={2008},
  organization={IEEE}
}

@article{kamiya2002ccfinder,
  title={{CCFinder: A Multilinguistic Token-Based Code Clone Detection System for Large Scale Source Code}},
  author={Kamiya, Toshihiro and Kusumoto, Shinji and Inoue, Katsuro},
  journal={IEEE transactions on software engineering},
  volume={28},
  number={7},
  pages={654--670},
  year={2002},
  publisher={IEEE}
}

@inproceedings{sajnani2016sourcerercc,
  title={{SourcererCC: Scaling Code Clone Detection to Big-code}},
  author={Sajnani, Hitesh and Saini, Vaibhav and Svajlenko, Jeffrey and Roy, Chanchal K and Lopes, Cristina V},
  booktitle={Proceedings of the 38th international conference on software engineering},
  pages={1157--1168},
  year={2016}
}

@inproceedings{choi2025tiver,
  title={{TIVER: Identifying Adaptive Versions of C/C++ Third-Party Open-Source Components Using a Code Clustering Technique}},
  author={Choi, Youngjae and Woo, Seunghoon},
  booktitle={Proceedings of the 47th International Conference on Software Engineering (ICSE). IEEE},
  year={2025},
  doi = {10.1109/ICSE55347.2025.00188}
}

@inproceedings{wu2023ossfp,
  title={{OSSFP: Precise and Scalable C/C++ Third-Party Library Detection using Fingerprinting Functions}},
  author={Wu, Jiahui and Xu, Zhengzi and Tang, Wei and Zhang, Lyuye and Wu, Yueming and Liu, Chengyue and Sun, Kairan and Zhao, Lida and Liu, Yang},
  booktitle={2023 IEEE/ACM 45th International Conference on Software Engineering (ICSE)},
  pages={270--282},
  year={2023},
  organization={IEEE}
}

@inproceedings{duan2017identifying,
  title={{Identifying Open-Source License Violation and 1-day Security Risk at Large Scale}},
  author={Duan, Ruian and Bijlani, Ashish and Xu, Meng and Kim, Taesoo and Lee, Wenke},
  booktitle={Proceedings of the 2017 ACM SIGSAC Conference on Computer and Communications Security},
  pages={2169--2185},
  year={2017}
}

@inproceedings{jiang2023third,
  title={{Third-Party Library Dependency for Large-Scale SCA in the C/C++ Ecosystem: How Far Are We?}},
  author={Jiang, Ling and Yuan, Hengchen and Tang, Qiyi and Nie, Sen and Wu, Shi and Zhang, Yuqun},
  booktitle={Proceedings of the 32nd ACM SIGSOFT International Symposium on Software Testing and Analysis},
  pages={1383--1395},
  year={2023},
  doi = {10.1145/3597926.3598143}
}

@inproceedings{woo2021centris,
  title={{CENTRIS: A Precise and Scalable Approach for Identifying Modified Open-Source Software Reuse}},
  author={Woo, Seunghoon and Park, Sunghan and Kim, Seulbae and Lee, Heejo and Oh, Hakjoo},
  booktitle={2021 IEEE/ACM 43rd International Conference on Software Engineering (ICSE)},
  pages={860--872},
  year={2021},
  organization={IEEE},
  doi = {10.1109/ICSE43902.2021.00083}
}

@inproceedings{shan2023gitor,
  title={{Gitor: Scalable Code Clone Detection by Building Global Sample Graph}},
  author={Shan, Junjie and Dou, Shihan and Wu, Yueming and Wu, Hairu and Liu, Yang},
  booktitle={Proceedings of the 31st ACM Joint European Software Engineering Conference and Symposium on the Foundations of Software Engineering},
  pages={784--795},
  year={2023},
  doi = {10.1145/3611643.3616371}
}

@inproceedings{gode2009incremental,
  title={{Incremental Clone Detection}},
  author={G{\"o}de, Nils and Koschke, Rainer},
  booktitle={2009 13th European conference on software maintenance and reengineering},
  pages={219--228},
  year={2009},
  organization={IEEE}
}

\end{document}